\documentclass{livrev}
\usepackage{amsfonts}
\usepackage{latexsym}

\begin{document}

\title{Initial Data for Numerical Relativity}

\author{Gregory B.\ Cook \\
	Department of Physics, Wake Forest University \\
	Winston-Salem, North Carolina\ \ 27109-7507 \\
	{\tt e-mail:cookgb@wfu.edu} \\
	{\tt http://www.wfu.edu/{\protect\~\protect}cookgb/}
	}

\date{}
\maketitle

\begin{abstract}
Initial data are the starting point for any numerical simulation.  In
the case of numerical relativity, Einstein's equations constrain our
choices of these initial data.  We will examine several of the
formalisms used for specifying Cauchy initial data in the $3\!+\!1$
decomposition of Einstein's equations.  We will then explore how these
formalisms have been used in constructing initial data for spacetimes
containing black holes and neutron stars.  In the topics discussed,
emphasis is placed on those issues that are important for obtaining
astrophysically realistic initial data for compact binary coalescence.
\end{abstract}


\newpage

\section{Introduction}\label{sec:introduction}

The goal of numerical relativity is to study spacetimes that cannot be
studied by analytic means.  The focus is therefore primarily on
dynamical systems.  Numerical relativity has been applied in many
areas: cosmological models, critical phenomena, perturbed black holes
and neutron stars, and the coalescence of black holes and neutron
stars, for example.  In any of these cases, Einstein's equations can
be formulated in several ways that allow us to evolve the dynamics.
While Cauchy methods have received a majority of the attention,
characteristic and Reggi calculus based methods have also been used.
All of these methods begin with a snapshot of the gravitational fields
on some hypersurface, \emph{the initial data}, and evolve these data
to neighboring hypersurfaces.

The focus of this review is on the initial data needed for Cauchy
evolutions of Einstein's equations.  These initial data cannot be
freely specified in their entirety.  Rather they are subject to
certain constraints which must be satisfied.  Because of the
nonlinearity of Einstein's equations, there is no unique way of
choosing which pieces of the initial data can be freely specified and
which are constrained.  In \S\ref{sec:init-value-equat}, I will look
at the various formalisms that exist for expressing the initial-value
equations.  I will use the $3\!+\!1$ (or ADM) decomposition of Einstein's
equations throughout and I begin the review with a brief introduction
of this in \S\ref{sec:initial-data}.  In \S\ref{sec:york-lichn-conf},
\S\ref{sec:thin-sandw-decomp}, and \S\ref{sec:stationary-solutions}, I
will explore the most important and widely used decompositions that
have been developed.

In the remainder of the review, I will focus on initial data for black
holes and neutron stars.  Section~\ref{sec:black-hole-initial} deals
with black-hole initial data, which has received considerable
attention over the years.  The evolution of black-hole spacetimes is
of particular importance because it allows for the study of pure
geometrodynamics.  The spacetime is either pure vacuum or any matter
can be hidden behind an event horizon.  Thus, black-hole spacetimes
allow us to study the dynamics of gravity alone.  In
\S\ref{sec:classic-solutions}, I begin with a review of some of the
classic analytic solutions of the initial-data equations.
Section~\ref{sec:general-multi-hole} covers current methods for
constructing general multi-black-hole initial data sets and also
explores some of the limitations of these data sets.
Section~\ref{sec:exact-solutions} deals with a class of black-hole
initial data that has received little attention until recently and
which I feel may play an important role in future work on constructing
black-hole initial data.  Finally, \S\ref{sec:quasicirc-binary}
examines the issue of constructing black-hole initial data for
quasiequilibrium binaries.

This last topic is of extreme importance.  Several laser
interferometer gravitational wave detectors will become operational in
the near future and the coalescence of black-hole binaries is
considered to be one of the strongest candidates for detection by the
earliest generation of detectors.  The chances of detecting these
events and then unraveling the information contained in the
gravitational wave signals will be greatly increased if we have
accurate numerical simulations of black-hole binary coalescence.
While post-Newtonian techniques can be used to simulate the inspiral
of a compact binary system, the final plunge and coalescence must be
simulated numerically.  Astrophysically realistic initial data will be
needed before these simulations can provide reliable results.

The final plunge and coalescence of neutron-star binaries must also be
simulated numerically.  Section~\ref{sec:neutron-star-initial}
examines the construction of neutron-star initial data.  The major
difference between black-hole and neutron-star initial data is the
need to deal with the neutron star's matter.  In this section, I will
deal with the neutron star matter in general, without considering any
particular equation of state.  In \S\ref{sec:hydr-equil}, I look at
the issue of hydrostatic equilibrium of matter.
Section~\ref{sec:isol-neutr-stars} takes a brief look at constructing
equilibrium initial data for isolated neutron stars.  Finally,
\S\ref{sec:neutr-star-binar} examines the issues involved in
constructing quasiequilibrium neutron-star binary initial data.  In
particular, we look at recent advances in constructing irrotational
fluid models.

This article is far from a complete review of all the work that has
been done on initial data for numerical relativity.  I offer my
apologies to all whose work I have not included.  In particular, I
have not covered any of the issues associated with existence and
uniqueness of solutions to the constraint equations, nor dealt
extensively with the issue of asymptotic falloff rates.  I welcome
correspondence on topics that you feel should be covered in this
review, references that you think should be included, and the
inevitable typographical errors.

\subsection{Conventions}
\label{sec:conventions}

I will use a 4-metric $g_{\mu\nu}$ with signature $(-+++)$.  Following
the MTW~\cite{MTW} conventions, I define the Riemann tensor as
\begin{equation}\label{eq:Riemann}
	R^\mu{}_{\nu\alpha\beta} \equiv
		  \partial_\alpha\Gamma^\mu{}_{\nu\beta}
		- \partial_\beta\Gamma^\mu{}_{\nu\alpha}
		+ \Gamma^\mu{}_{\sigma\alpha}\Gamma^\sigma{}_{\nu\beta}
		- \Gamma^\mu{}_{\sigma\beta}\Gamma^\sigma{}_{\nu\alpha}.
\end{equation}
The Ricci tensor is defined as
\begin{equation}\label{eq:Ricci}
	R_{\mu\nu} \equiv R^\sigma{}_{\mu\sigma\nu},
\end{equation}
and the Einstein tensor as
\begin{equation}\label{eq:Einstein}
	G_{\mu\nu} \equiv R_{\mu\nu} - {\textstyle\frac12}g_{\mu\nu}R.
\end{equation}
A spatial 3-metric will be written as $\gamma_{ij}$ and the Riemann
and Ricci tensors associated with it will be defined by
(\ref{eq:Riemann}) and (\ref{eq:Ricci}).  Four-dimensional indices
will be denoted by Greek letters, while 3-dimensional indices will be
denoted by Latin letters.

To avoid confusion, the covariant derivative and Ricci tensor
associated with $\gamma_{ij}$ will be written with
over-bars---$\bar\nabla_j$ and $\bar{R}_{ij}$.  I will also
frequently deal with an auxiliary 3-dimensional space with a metric
that is conformally related to the metric $\gamma_{ij}$ of the
physical space.  The metric for this space will be denoted
$\tilde\gamma_{ij}$.  The covariant derivative and Ricci tensor
associated with this metric will be written with
tildes---$\tilde\nabla_j$ and $\tilde{R}_{ij}$.  Other quantities that
have a conformal relationship to quantities in the physical space will
also be written with a tilde over them.

\newpage

\section{The Initial-Value Equations}
\label{sec:init-value-equat}

Einstein's equations, $G_{\mu\nu} = 8\pi G T_{\mu\nu}$, represent ten
independent equations.  Since there are ten equations and ten
independent components of the four-metric $g_{\mu\nu}$, it seems that
we have the same number of equations as unknowns.  From the definition
of the Einstein tensor (\ref{eq:Einstein}), we see that these ten
equations are linear in the second derivatives and quadratic in the
first derivatives of the metric.  We might expect that these ten
second-order equations represent evolution equations for the ten
components of the metric.  However, a close inspection of the
equations reveals that only six of the ten involve second
time-derivatives of the metric.  The remaining four equations are not
evolution equations.  Instead, they are constraint equations.  The
full system of equations is still well posed, however, because of the
Bianchi identities
\begin{equation}\label{eq:Bianchi}
	\nabla_\nu G^{\mu\nu} \equiv 0.
\end{equation}
The four constraint equations appear as a result of the general
covariance of Einstein's theory, which gives us the freedom to apply
general coordinate transformations to each of the four coordinates
and leave the interval,
\begin{equation}\label{eq:interval}
	{\rm d}s^2 = g_{\mu\nu} {\rm d}x^\mu {\rm d}x^\nu,
\end{equation}
unchanged.

If we consider Einstein's equations as a Cauchy problem\footnote{We
could also formulate Einstein's equations as a characteristic
initial-value problem, but we will not pursue that approach in this
paper.}, we find that the ten equations separate into a set of four
constraint or initial-value equations, and six evolution or
dynamical equations.  If the four initial-value equations are
satisfied on some spacelike hypersurface, which we can label with
$t=0$, then the Bianchi identities (\ref{eq:Bianchi}) guarantee that
the evolution equations preserve the constraints on neighboring
spacelike hypersurfaces.

\subsection{Initial Data}
\label{sec:initial-data}

In the Cauchy formulation of Einstein's equations, we begin by
foliating the 4-dimensional manifold as a set of spacelike,
3-dimensional hypersurfaces (or slices) $\{\Sigma\}$.  These slices
are labeled by a parameter $t$ or, more simply, each slice of the
4-dimensional manifold is a $t=\mbox{constant}$ hypersurface.
Following the standard $3\!+\!1$
decomposition~\cite{ADM,york-Sources,choquet-york-GRG}, we let $n^\mu$
be the future-pointing timelike unit normal to the slice, with
\begin{equation}\label{eq:unit-normal}
	n^\mu \equiv -\alpha\nabla^\mu t.
\end{equation}
Here, $\alpha$ is called the \emph{lapse function} (frequently denoted
$N$ in the literature).  The scalar lapse function sets the proper
interval measured by observers as they move between slices on a path
that is normal to the hypersurface (so-called normal observers):
\begin{equation}\label{eq:lapse-interval}
	{\rm d}s|_{\mbox{\small along $n^\mu$}} = \alpha {\rm d}t.
\end{equation}
Of course, there is no reason that observers must move along a path
normal to the hypersurface.  In general, we can define the time vector
as
\begin{equation}\label{eq:time}
	t^\mu \equiv \alpha n^\mu + \beta^\mu,
\end{equation}
where
\begin{equation}\label{eq:shift-const}
	\beta^\mu n_\mu \equiv 0.
\end{equation}
Here, $\beta^\mu$ is called the \emph{shift vector} (frequently
denoted $N^\mu$ in the literature).

Because of (\ref{eq:shift-const}), $\beta^\mu$ has only three
independent components and is a \emph{spatial} vector, tangent to the
hypersurface on which it resides.  At this point, it is convenient to
introduce a coordinate system adapted to the foliation $\{\Sigma\}$.
Let $x^i$ be the spatial coordinates in the slice.  The fourth
coordinate, $t$, is the parameter labeling each slice.  With this
adapted coordinate system, we find that 3-dimensional coordinate
values remain constant as we move between slices along the $t^\mu$
direction (\ref{eq:time}).  The four parameters, $\alpha$ and
$\beta^i$, are a manifestation of the 4-dimensional coordinate
invariance, or gauge freedom, in Einstein's theory.  If we let
$\gamma_{ij}$ represent the metric of the spacelike hypersurfaces,
then we can rewrite the interval (\ref{eq:interval}) as
\begin{equation}\label{eq:3+1-interval}
	{\rm d}s^2 = -\alpha^2{\rm d}t^2
		+ \gamma_{ij}({\rm d}x^i + \beta^i{\rm d}t)
				({\rm d}x^j + \beta^j{\rm d}t).
\end{equation}

In the Cauchy formulation of Einstein's equations, $\gamma_{ij}$ is
regarded as the fundamental variable and values for its components
must be given as part of a well-posed initial-value problem.  Since
Einstein's equations are second order, we must also specify something
like a time derivative of the metric.  For this, we use the second
fundamental form, or extrinsic curvature, of the slice, $K_{ij}$,
defined\footnote{A different sign choice for defining the extrinsic
curvature is sometimes found in the literature.} by
\begin{equation}\label{eq:extr-curv}
	K_{ij} \equiv -{\textstyle\frac12}{\cal L}_n\gamma_{ij},
\end{equation}
where ${\cal L}_n$ denotes the Lie derivative along the $n^\mu$ 
direction.

Together, $\gamma_{ij}$ and $K_{ij}$ are the minimal set of initial
data that must be specified for a Cauchy evolution of Einstein's
equations.  The metric $\gamma_{ij}$ on a hypersurface is induced on
that surface by the 4-metric $g_{\mu\nu}$.  This means that the values
$\gamma_{ij}$ receives depend on how $\Sigma$ is embedded in the full
spacetime.  In order for the foliation of slices $\{\Sigma\}$ to fit
into the higher-dimensional space, they must satisfy the
Gauss--Codazzi--Ricci conditions.  Combining these conditions with
Einstein's equations, and using (\ref{eq:3+1-interval}), the six
evolution equations become
\begin{eqnarray}\label{eq:K-evolution}
	\partial_t K_{ij} &=& \alpha \left[
		\bar{R}_{ij} - 2K_{i\ell}K^\ell_j + K K_{ij}
		- 8\pi G S_{ij} + 4\pi G\gamma_{ij}(S - \rho)\right] \\
		&& \mbox{} 
		- \bar\nabla_i\bar\nabla_j\alpha
		+ \beta^\ell\bar\nabla_\ell K_{ij}
		+ K_{i\ell}\bar\nabla_j\beta^\ell
		+ K_{j\ell}\bar\nabla_i\beta^\ell. \nonumber
\end{eqnarray}
Here, $\bar\nabla_i$ is the spatial covariant derivative compatible
with $\gamma_{ij}$, $\bar{R}_{ij}$ is the Ricci tensor associated with
$\gamma_{ij}$, $K\equiv K^i_i$, $\rho$ is the matter energy density,
$S_{ij}$ is the matter stress tensor, and $S\equiv S^i_i$.\footnote{
The stress-energy tensor is decomposed as $T_{\mu\nu} = S_{\mu\nu} +
n_\mu j_\nu + n_\nu j_\mu + n_\mu n_\nu\rho$ or, equivalently, $S_{ij}
\equiv T_{ij}$, $j_i \equiv -n^\mu T_{i\mu}$, and $\rho \equiv n^\mu
n^\nu T_{\mu\nu}$.  Note that the matter terms in
Eqs.~(\ref{eq:K-evolution}), (\ref{eq:Hamiltonian-const}), and
(\ref{eq:momentum-const}) are defined with respect to a normal
observer.  This is in contrast to the usual definition of $\rho$ with
respect to the rest frame of the fluid in hydrodynamics.}  We have
also used the fact that in our adapted coordinate system, ${\cal L}_t
\equiv \partial_t$.  The set of second-order evolution equations is
completed by rewriting the definition of the extrinsic curvature
(\ref{eq:extr-curv}) as
\begin{equation}\label{eq:g-evolution}
	\partial_t \gamma_{ij} = -2\alpha K_{ij}
		+ \bar\nabla_i\beta_j
		+ \bar\nabla_j\beta_i.
\end{equation}

Equations~(\ref{eq:K-evolution}) and (\ref{eq:g-evolution}) are a
first-order representation of a complete set of evolution equations
for given initial data $\gamma_{ij}$ and $K_{ij}$.  However, the data
cannot be freely specified in their entirety.  The four constraint
equations, following the same procedure outlined above for the
evolution equations, become
\begin{equation}\label{eq:Hamiltonian-const}
	\bar{R} + K^2 - K_{ij}K^{ij} = 16\pi G\rho
\end{equation}
and
\begin{equation}\label{eq:momentum-const}
	\bar\nabla_j\left(K^{ij} - \gamma^{ij}K\right) = 8\pi G j^i.
\end{equation}
Here, $\bar{R}\equiv\bar{R}^i_i$ and $j^i$ is the matter momentum
density.  Equation~(\ref{eq:Hamiltonian-const}) is referred to as the
Hamiltonian or scalar constraint, while (\ref{eq:momentum-const}) are
referred to as the momentum or vector constraints.  Valid initial data
for the evolution equations (\ref{eq:K-evolution}) and
(\ref{eq:g-evolution}) must satisfy this set of constraints.  And, as
mentioned earlier, the Bianchi identities (\ref{eq:Bianchi}) guarantee
that the evolution equations will preserve the constraints on future
slices of the evolution.

As we will see below, the Hamiltonian constraint
(\ref{eq:Hamiltonian-const}) most naturally constrains the 3-metric
$\gamma_{ij}$, while the momentum constraints
(\ref{eq:momentum-const}) naturally constrain the extrinsic curvature
$K_{ij}$.  Taking the constraints into consideration, it seems that
the 3-metric has five degrees of freedom remaining, while the
extrinsic curvature has three.  But we know that the gravitational
field in Einstein's theory has two dynamical degrees of freedom, so we
expect that both $\gamma_{ij}$ and $K_{ij}$ should each have only two
free components.  The answer to this problem is, once again, the
coordinate invariance of Einstein's theory.  This seems strange at
first, because we have already used the coordinate invariance of the
theory to narrow our scope from the ten components of the 4-metric to
the six components of the 3-metric.  However, the lapse and shift do
not completely specify the coordinate gauge.  Rather, they specify how
an initial choice of gauge will evolve with the foliation.  The metric
on a given hypersurface retains full 3-dimensional coordinate
invariance, reducing the number of freely specifiable components to
two~\cite{york-1971,york-1972}.  There also remains one degree of
gauge freedom associated with the time coordinate which must be fixed.
Each hypersurface represents a $t=\mbox{const.}$ slice of the
spacetime, so how the initial hypersurface is embedded in the full
4-dimensional manifold represents our temporal gauge choice.  There is
no unique way to specify this choice, but it is often convenient to
let the trace of the extrinsic curvature $K$ represent this temporal
gauge choice~\cite{york-1972}.  Thus, we find that we \emph{are}
allowed to choose freely five components of the 3-metric and three
components of the extrinsic curvature.  However, only \emph{two} of
the components for each field represent dynamical degrees of freedom,
the remainder are gauge degrees of freedom.

The four constraint equations, (\ref{eq:Hamiltonian-const}) and
(\ref{eq:momentum-const}), represent conditions which the 3-metric and
extrinsic curvature must satisfy.  But, they do not specify which
components (or combination of components) are constrained and which
are freely specifiable.  In the weak field limit where Einstein's
equations can be linearized, there are clear ways to determine which
components are dynamic, which are constrained, and which are gauge.
However, in the full nonlinear theory, there is no unique
decomposition.  In this case, one must choose a method for decomposing
the constraint equations.  The goal is to transform the equations into
standard elliptic forms which can be solved given appropriate boundary
conditions~\cite{omurchada-york-1973,omurchada-york-1974-II,york-Sources,Essays-in-GR}.
Each different decomposition yields a unique set of elliptic equations
to be solved \emph{and} a unique set of freely specifiable parameters
which must be fixed somehow.  Seemingly similar sets of assumptions
applied to different decompositions can lead to physically different
initial conditions.

\subsection{York--Lichnerowicz Conformal Decompositions}
\label{sec:york-lichn-conf}

For general initial-data configurations, the most widely used
class of constraint decompositions are the York--Lichnerowicz
conformal decompositions.  At their heart are a conformal
decomposition of the metric and certain components of the extrinsic
curvature, together with a transverse-traceless decomposition of the
extrinsic curvature.

First, the metric is decomposed into a conformal factor $\psi$
multiplying an auxiliary
3-metric~\cite{lichnerowicz-1944,york-1971,york-1972}:
\begin{equation}\label{eq:conformal-metric}
	\gamma_{ij} \equiv \psi^4\tilde\gamma_{ij}.
\end{equation}
The auxiliary 3-metric $\tilde\gamma_{ij}$ is often called the
conformal or background 3-metric, and it carries five degrees of
freedom.  Its natural definition is given by
\begin{equation}\label{eq:normalized-conf-metric}
	\tilde\gamma_{ij} = \gamma^{-\frac13}\gamma_{ij},
\end{equation}
leaving $\tilde\gamma = 1$, but we are free to choose any
normalization for $\tilde\gamma$.  Using (\ref{eq:conformal-metric}), we
can rewrite the Hamiltonian constraint (\ref{eq:Hamiltonian-const}) as
\begin{equation}\label{eq:conf-Hamiltonian-const-v1}
	\tilde\nabla^2\psi - {\textstyle\frac18}\psi\tilde{R}
		- {\textstyle\frac18}\psi^5K^2
		+ {\textstyle\frac18}\psi^5K_{ij}K^{ij} = -2\pi G\psi^5\rho,
\end{equation}
where $\tilde\nabla^2 \equiv \tilde\nabla^i\tilde\nabla_i$ is the
scalar Laplace operator, and $\tilde\nabla_i$ and $\tilde{R}$ are the
covariant derivative and Ricci scalar associated with
$\tilde\gamma_{ij}$.  Equation~(\ref{eq:conf-Hamiltonian-const-v1}) is
a quasilinear elliptic equation for the conformal factor $\psi$, and
we see that the Hamiltonian constraint naturally constrains the
3-metric.

The conformal decomposition of the Hamiltonian constraint was proposed
by Lichnerowicz.  But, the key to the full decomposition is the
treatment of the extrinsic curvature introduced by
York~\cite{york-1973,york-1974}.  This begins by splitting the
extrinsic curvature into its trace and tracefree parts,
\begin{equation}\label{eq:trace-free-K}
	K_{ij} \equiv A_{ij} + {\textstyle\frac13}\gamma_{ij}K.
\end{equation}
The decomposition proceeds by using the fact that we can covariantly
split any symmetric tracefree tensor as follows:
\begin{equation}\label{eq:gen-TT-decomp}
	{\cal S}^{ij} \equiv ({\mathbb L}X)^{ij} + T^{ij}.
\end{equation}
Here, $T^{ij}$ is a symmetric, transverse-traceless tensor (i.e.,
$\nabla_jT^{ij} = 0$ and $T^i_i=0$) and
\begin{equation}\label{eq:TT-op-def}
	({\mathbb L}X)^{ij} \equiv \nabla^iX^j +\nabla^jX^i
		- {\textstyle\frac23}\gamma^{ij}\nabla_\ell X^\ell.
\end{equation}
After separating out the transverse-traceless portion of ${\cal
S}^{ij}$, what remains, $({\mathbb L}X)^{ij}$, is referred to as its
``longitudinal'' part.  We now want to apply this transverse-traceless
decomposition to the tracefree part of the extrinsic curvature
$A_{ij}$.  However, the conformal decomposition of the metric leaves
us with at least two ways to proceed.

The goal of the decomposition is to produce a coupled set of elliptic
equations to be solved with some prescribed boundary conditions.  We
have already reduced the Hamiltonian constraint to an elliptic
equation being solved on a \emph{background} space in terms of
differential operators that are compatible with the conformal
3-metric.  In the end, we want to reduce the momentum constraints to
a set of elliptic equations based on differential operators that are
compatible with the same conformal 3-metric.  But, the longitudinal
operator (\ref{eq:TT-op-def}) can be defined with respect to
\emph{any} metric space.  In particular, it is natural to consider
decompositions with respect to both the physical and conformal
3-metrics.

\subsubsection{Conformal Transverse-Traceless Decomposition}
\label{sec:conf-transv-trac}

Let us first consider decomposing $A^{ij}$ with respect to the
conformal 3-metric~\cite{york-Sources,york-piran-StG}.  As we will see,
when certain assumptions are made, this decomposition has the
advantage of producing a simpler set of elliptic equations that must
be solved.  The first step is to define the conformal tracefree
extrinsic curvature $\tilde{A}^{ij}$ by
\begin{equation}\label{eq:conf-A}
	A^{ij} \equiv \psi^{-10}\tilde{A}^{ij}\qquad\mbox{or}\qquad
	A_{ij} \equiv \psi^{-2}\tilde{A}_{ij}.
\end{equation}
Next, the transverse-traceless decomposition is applied to the conformal
extrinsic curvature,
\begin{equation}\label{eq:decomp-back-A}
	\tilde{A}^{ij} \equiv (\tilde{\mathbb L}X)^{ij} + \tilde{Q}^{ij}.
\end{equation}
Note that the longitudinal operator $\tilde{\mathbb L}$ and the
symmetric, transverse-tracefree tensor $\tilde{Q}^{ij}$ are both
defined with respect to covariant derivatives compatible with
$\tilde\gamma_{ij}$.

Applying equations~(\ref{eq:conformal-metric}),
(\ref{eq:trace-free-K}), (\ref{eq:TT-op-def}), (\ref{eq:conf-A}), and
(\ref{eq:decomp-back-A}) to the momentum constraints
(\ref{eq:momentum-const}), we find that they simplify to
\begin{equation}\label{eq:conf-mom-v1}
	\tilde\Delta_{\mathbb L}X^i =
		{\textstyle\frac23}\psi^6\tilde\nabla^iK
		+ 8\pi G\psi^{10}j^i,
\end{equation}
where
\begin{equation}\label{eq:long-elliptic-op}
	\tilde\Delta_{\mathbb L}X^i \equiv
		\tilde\nabla_j({\mathbb L}X)^{ij} =
		\tilde\nabla^2X^i
		+ {\textstyle\frac13}\tilde\nabla^i(\tilde\nabla_jX^j)
		+ \tilde{R}^i_jX^j,
\end{equation}
and we have used the fact that
\begin{equation}\label{eq:conf-diverg-ident}
	\bar\nabla_j{\cal S}^{ij} =
		\psi^{-10}\tilde\nabla_j(\psi^{10}{\cal S}^{ij})
\end{equation}
for any symmetric tracefree tensor ${\cal S}^{ij}$.

In deriving equation~(\ref{eq:conf-mom-v1}), we have also used the fact
that $\tilde{Q}^{ij}$ is transverse (i.e.\
$\tilde\nabla_j\tilde{Q}^{ij}=0$).  However, in general, we will not
know if a given symmetric tracefree tensor, say $\tilde{M}^{ij}$, is
transverse.  By using (\ref{eq:gen-TT-decomp}) we can obtain its
transverse-traceless part $\tilde{Q}^{ij}$ via
\begin{equation}\label{eq:transverse-proc-v1}
	\tilde{Q}^{ij} \equiv \tilde{M}^{ij} - (\tilde{\mathbb L}Y)^{ij},
\end{equation}
and using the fact that if $\tilde{Q}^{ij}$ is transverse, we find
\begin{equation}\label{eq:trans-dec-eqn-v1}
	\tilde\nabla_j\tilde{Q}^{ij} \equiv 0 =
		\tilde\nabla_j\tilde{M}^{ij} - \tilde\Delta_{\mathbb L}Y^i.
\end{equation}
Thus, Eqs.~(\ref{eq:transverse-proc-v1}) and
(\ref{eq:trans-dec-eqn-v1}) give us a general way of
\emph{constructing} the required symmetric transverse-traceless tensor
from a general symmetric traceless tensor.

Using the linearity of $\tilde{\mathbb L}$, we can rewrite
(\ref{eq:decomp-back-A}) as
\begin{equation}\label{eq:decomp-back-A-gen}
	\tilde{A}^{ij} = (\tilde{\mathbb L}V)^{ij} + \tilde{M}^{ij},
\end{equation}
where
\begin{equation}\label{eq:vector-pot-gen-v1}
	V^i \equiv X^i - Y^i.
\end{equation}
Similarly, using the linearity of $\tilde\Delta_{\mathbb L}$, we can rewrite
(\ref{eq:conf-mom-v1}) as
\begin{equation}\label{eq:conf-mom-v2}
	\tilde\Delta_{\mathbb L}V^i =
		{\textstyle\frac23}\psi^6\tilde\nabla^iK
		- \tilde\nabla_j\tilde{M}^{ij} + 8\pi G\psi^{10}j^i.
\end{equation}
By solving directly for $V^i$, we can combine the steps of decomposing
$\tilde{M}^{ij}$ with that of solving the momentum constraints.

After applying (\ref{eq:trace-free-K}) and (\ref{eq:conf-A}) to the
Hamiltonian constraint (\ref{eq:conf-Hamiltonian-const-v1}), we obtain
the following full decomposition, which I will list together here for
convenience:
\begin{eqnarray}\label{eq:conf-YL-decomp}
	\gamma_{ij} &=& \psi^4\tilde\gamma_{ij} \nonumber\\
	K^{ij} &=& \psi^{-10}\tilde{A}^{ij}
		 + {\textstyle\frac13}\psi^{-4}\tilde\gamma^{ij}K \nonumber\\
	\tilde{A}^{ij} &=& (\tilde{\mathbb L}V)^{ij}+\tilde{M}^{ij}  \\
	\tilde\Delta_{\mathbb L}V^i
		- {\textstyle\frac23}\psi^6\tilde\nabla^iK &=&
		- \tilde\nabla_j\tilde{M}^{ij} + 8\pi G\psi^{10}j^i \nonumber\\
	\tilde\nabla^2\psi - {\textstyle\frac18}\psi\tilde{R}
		- {\textstyle\frac1{12}}\psi^5K^2
		+ {\textstyle\frac18}\psi^{-7}\tilde{A}_{ij}\tilde{A}^{ij}
		&=& -2\pi G\psi^5\rho \nonumber
\end{eqnarray}

In the decomposition given by (\ref{eq:conf-YL-decomp}), we are free
to specify a symmetric tensor $\tilde\gamma_{ij}$ as the conformal
3-metric, a symmetric tracefree tensor $\tilde{M}^{ij}$, and a scalar
function $K$.  Then, with given matter energy and momentum densities,
$\rho$ and $j^i$, and appropriate boundary conditions, the coupled set
of constraint equations for $\psi$ and $V^i$ are solved.  Finally,
given the solutions, we can construct the physical initial data,
$\gamma_{ij}$ and $K^{ij}$.

The decomposition outlined above has the interesting property that if
we choose $K$ to be constant and if the momentum density
vanishes\footnote{Or, if we solve the momentum constraint in terms of
a background momentum density $\tilde{j}^i \equiv \psi^{10}j^i$, the
momentum constraint decouples with nonvanishing $\tilde{j}^i$.}, then
the momentum constraint equations fully decouple from the Hamiltonian
constraint.  As we will see later, this simplification has proven to
be useful.

\subsubsection{Physical Transverse-Traceless Decomposition}
\label{sec:phys-transv-trac}

Alternatively, we can decompose $A^{ij}$ with respect to the physical
3-metric~\cite{omurchada-york-1974-I,omurchada-york-1974-II,omurchada-york-1976}.
We decompose the extrinsic curvature as
\begin{equation}\label{eq:decomp-phys-A}
	A^{ij} \equiv (\bar{\mathbb L}W)^{ij} + Q^{ij}.
\end{equation}
In this case, the longitudinal operator $\bar{\mathbb L}$ and the
symmetric transverse-tracefree tensor $Q^{ij}$ are both defined with
respect to covariant derivatives compatible with $\gamma_{ij}$.

Applying equations~(\ref{eq:conformal-metric}),
(\ref{eq:trace-free-K}), (\ref{eq:TT-op-def}),
(\ref{eq:decomp-phys-A}), and (\ref{eq:conf-diverg-ident}) to the
momentum constraint (\ref{eq:momentum-const}), we find that it
simplifies to
\begin{equation}\label{eq:phys-mom-v1}
	\tilde\Delta_{\mathbb L}W^i
		+ 6(\tilde{\mathbb L}W)^{ij}\tilde\nabla_j\ln\psi =
		{\textstyle\frac23}\tilde\nabla^iK
		+ 8\pi G\psi^4j^i,
\end{equation}
where we have used the fact that
\begin{equation}\label{eq:long-op-conf-ident}
	(\bar{\mathbb L}W)^{ij} = \psi^{-4}(\tilde{\mathbb L}W)^{ij}.
\end{equation}

As in the previous section, we will obtain the symmetric
transverse-traceless tensor $Q^{ij}$ from a general symmetric
tracefree tensor $\tilde{M}^{ij}$ by using (\ref{eq:gen-TT-decomp}).
In this case, we take
\begin{equation}\label{eq:transverse-proc-v2}
	Q^{ij} \equiv \psi^{-10}\tilde{M}^{ij} - (\bar{\mathbb L}Z)^{ij},
\end{equation}
and use the fact that $Q^{ij}$ is transverse, to obtain
\begin{equation}\label{eq:trans-dec-eqn-v2}
	\tilde\Delta_{\mathbb L}Z^i
		+ 6(\tilde{\mathbb L}Z)^{ij}\tilde\nabla_j\ln\psi =
		\psi^{-6}\tilde\nabla_j\tilde{M}^{ij}.
\end{equation}
Again, we can define
\begin{equation}\label{eq:vector-pot-gen-v2}
	V^i \equiv W^i - Z^i,
\end{equation}
and use the linearity of $\tilde{\mathbb L}$ and
$\tilde\Delta_{\mathbb L}$ to combine the process of obtaining the
transverse-traceless part of $M^{ij}$ and solving the momentum
constraints.  We obtain the following full decomposition, which I will
list together here for convenience\footnote{Note that in
(\ref{eq:phys-YL-decomp}) we have \emph{not} used the usual conformal
scaling of the extrinsic curvature given in
equation~(\ref{eq:conf-A}).}:
\begin{eqnarray}\label{eq:phys-YL-decomp}
	\gamma_{ij} &=& \psi^4\tilde\gamma_{ij} \nonumber\\
	K^{ij} &=& \psi^{-4}\left(\tilde{A}^{ij}
		 + {\textstyle\frac13}\tilde\gamma^{ij}K\right) \nonumber\\
	\tilde{A}^{ij} &=& (\tilde{\mathbb L}V)^{ij}+\psi^{-6}\tilde{M}^{ij} \\
	\tilde\Delta_{\mathbb L}V^i
		+ 6(\tilde{\mathbb L}V)^{ij}\tilde\nabla_j\ln\psi &=&
		{\textstyle\frac23}\tilde\nabla^iK
		- \psi^{-6}\tilde\nabla_j\tilde{M}^{ij}
		+ 8\pi G\psi^4j^i \nonumber\\
	\tilde\nabla^2\psi - {\textstyle\frac18}\psi\tilde{R}
		- {\textstyle\frac1{12}}\psi^5K^2
		+ {\textstyle\frac18}\psi^5\tilde{A}_{ij}\tilde{A}^{ij}
		&=& -2\pi G\psi^5\rho \nonumber
\end{eqnarray}

In the decomposition given by (\ref{eq:phys-YL-decomp}), we are again
free to specify a symmetric tensor $\tilde\gamma_{ij}$ as the
conformal 3-metric, a symmetric tracefree tensor $\tilde{M}^{ij}$, and
a scalar function $K$.  Then, with given matter energy and momentum
densities, $\rho$ and $j^i$, and appropriate boundary conditions, the
coupled set of constraint equations for $\psi$ and $V^i$ are solved.
Finally, given the solutions, we can construct the physical initial
data, $\gamma_{ij}$ and $K^{ij}$.

Notice that, while very similar to the decomposition from
\S\ref{sec:conf-transv-trac}, the sets of equations are distinctly
different.  In general, if we make the same choices for the freely
specifiable data in both decompositions (i.e., we choose $\tilde\gamma_{ij}$,
$\tilde{M}^{ij}$, and $K$ the same), we will produce two different
sets of initial data.  Both will be equally valid solutions of the
constraint equations, but they will have distinct physical properties.

There is at least one exception to this.  Assume we have a valid set
of initial data $\gamma_{ij}$ and $K_{ij}$, which satisfies the
constraint equations (\ref{eq:Hamiltonian-const}) and
(\ref{eq:momentum-const}).  For any everywhere-positive function $\Psi$,
we define our freely specifiable data as follows:
\begin{eqnarray}\label{eq:conf-equiv-data}
	\tilde\gamma_{ij} &\equiv& \Psi^{-4}\gamma_{ij} \nonumber\\
	\tilde{M}^{ij} &\equiv& \Psi^{10}\left(K^{ij}
			- {\textstyle\frac13}\gamma^{ij}K\right) \\
	K &\equiv& K^i_i. \nonumber
\end{eqnarray}
Then the solution to both sets of equations, assuming we use correct
boundary conditions, will be $\psi = \Psi$ and $V^i=0$, which yields
the original data as the solution for each decomposition.

\subsection{Thin-Sandwich Decomposition}
\label{sec:thin-sandw-decomp}

The two general initial-value decompositions outlined in
\S\ref{sec:conf-transv-trac} and \S\ref{sec:phys-transv-trac}
require identical freely specified data ($\tilde\gamma_{ij}$,
$\tilde{M}^{ij}$, and $K$), yet they usually produce different
physical initial data.  One shortcoming of these approaches is that
they provide no direct insight into how to choose the freely
specifiable data.  All of the data are determined by schemes that
involve only a single spacelike hypersurface.  The resulting
constraint equations are independent of the kinematical variables
$\alpha$ and $\beta^i$ that govern how the coordinates move through
spacetime, and thus there is no connection to dynamics.  York's
thin-sandwich decomposition~\cite{york-1999} takes a different approach
by considering the evolution of the metric between two neighboring
hypersurfaces (the thin sandwich).  This decomposition is very similar
to an approach originated by
Wilson~\cite{wilson-mathews-Frontiers,cook-etal-1996}, but is somewhat
more general.  Perhaps the most attractive feature of this
decomposition is the insight it yields into the choice of the freely
specifiable data.

The decomposition begins with the standard conformal decomposition of
the 3-metric (\ref{eq:conformal-metric}).  However, we next make use
of the evolution equation for the metric (\ref{eq:g-evolution}) in
order to connect the 3-metrics on the two neighboring hypersurfaces.
Label the two slices by $t$ and $t^\prime$, with $t^\prime = t +
\delta{t}$, then $\gamma^\prime_{ij} = \gamma_{ij} +
\left(\partial_t\gamma_{ij}\right)\delta{t}$.  We would like to
specify how the 3-metric evolves, but we do not have full freedom to
do this.  We know we can freely specify only the conformal 3-metric,
and similarly, we are free to specify only the evolution of the
conformal 3-metric.  We make the following definitions:
\begin{equation}\label{eq:TS-u-def}
	u_{ij} \equiv \gamma^{\frac13}\partial_t(\gamma^{-\frac13}\gamma_{ij}),
\end{equation}
\begin{equation}\label{eq:TS-conf-u-def}
	\tilde{u}_{ij} \equiv \partial_t\tilde\gamma_{ij},
\end{equation}
and
\begin{equation}\label{eq:TS-udot-constraint}
	\tilde\gamma^{ij}\tilde{u}_{ij} \equiv 0.
\end{equation}
The latter definition is made for convenience, so that we can treat
$\psi$, $\tilde\gamma_{ij}$, and $\tilde{u}_{ij}$ as regular scalars
and tensors instead of as scalar- and tensor-densities within this
thin-sandwich formalism.

The conformal scaling of $u_{ij}$ follows directly from
(\ref{eq:conformal-metric}), (\ref{eq:TS-u-def}),
(\ref{eq:TS-conf-u-def}), (\ref{eq:TS-udot-constraint}), and the
identity that, for any small perturbation, $\delta\ln\gamma =
\gamma^{ij}\delta\gamma_{ij}$.  The result is
\begin{equation}\label{eq:TS-conf-u-scaling}
	u_{ij} = \psi^4\tilde{u}_{ij},
\end{equation}
which relies on the useful intermediate result that\footnote{
Equations~(\ref{eq:TS-udot-constraint}) and
(\ref{eq:TS-dt-det-conf-met}) take the place of defining
$\tilde\gamma=1$ and the subsequent necessity of treating $\psi$ and
$\tilde\gamma_{ij}$ as densities.}
\begin{equation}\label{eq:TS-dt-det-conf-met}
	\partial_t\tilde\gamma = 0.
\end{equation}

Equation~(\ref{eq:TS-u-def}) represents the tracefree part of the
evolution of the 3-metric, so (\ref{eq:g-evolution}) becomes
\begin{equation}\label{eq:TS-TF-gdot}
	u^{ij} = -2\alpha A^{ij} + (\bar{\mathbb L}\beta)^{ij}.
\end{equation}
Using the conformal scalings (\ref{eq:conf-A}),
(\ref{eq:long-op-conf-ident}), and (\ref{eq:TS-conf-u-scaling}), we
obtain
\begin{equation}\label{eq:TS-conf-gdot}
	\tilde{A}^{ij} = \frac{\psi^6}{2\alpha}\left(
		(\tilde{\mathbb L}\beta)^{ij} - \tilde{u}^{ij}\right).
\end{equation}

York has pointed out that it is natural to use the following conformal
rescaling of the lapse:
\begin{equation}\label{eq:conf-lapse-def}
	\alpha = \psi^6\tilde\alpha.
\end{equation}
This rescaling follows naturally from the ``slicing function'' that
replaces the usual lapse ($\alpha = \sqrt\gamma\tilde\alpha$) which
has been critical in solving several
problems~\cite{anderson-york-1998}.  It also results in the natural
conformal scaling (\ref{eq:conf-A}) postulated for the tracefree part
of the extrinsic curvature.  Substituting (\ref{eq:conf-lapse-def})
into (\ref{eq:TS-conf-gdot}) yields what is taken as the definition of
the tracefree part of the conformal extrinsic curvature,
\begin{equation}\label{eq:TS-conf-A-def}
	\tilde{A}^{ij} \equiv \frac1{2\tilde\alpha}\left(
		(\tilde{\mathbb L}\beta)^{ij} - \tilde{u}^{ij}\right).
\end{equation}

Because the tracefree extrinsic curvature satisfies the normal
conformal scaling, the Hamiltonian constraint will take on the same
form as in (\ref{eq:conf-YL-decomp}).  However, the momentum
constraint will have a very different form.  Combining
equations~(\ref{eq:conformal-metric}), (\ref{eq:trace-free-K}),
(\ref{eq:TT-op-def}), (\ref{eq:conf-A}), and (\ref{eq:TS-conf-A-def})
with the momentum constraint (\ref{eq:momentum-const}), we find that
it simplifies to
\begin{equation}\label{eq:TS-momentum-const}
	\tilde\Delta_{\mathbb L}\beta^i
		- (\tilde{\mathbb L}\beta)^{ij}\tilde\nabla_j\ln\tilde\alpha =
		- {\textstyle\frac43}\tilde\alpha\psi^6\tilde\nabla^iK
		+ \tilde\alpha\tilde\nabla\left(
			{\textstyle\frac1{\tilde\alpha}}\tilde{u}^{ij}\right)
		+ 16\pi G\tilde\alpha\psi^{10}j^i.
\end{equation}

Let us, for convenience, group together all the equations that
constitute the thin-sandwich decomposition:
\begin{eqnarray}\label{eq:TS-decomp}
	\gamma_{ij} &=& \psi^4\tilde\gamma_{ij} \nonumber\\
	K^{ij} &=& \psi^{-10}\tilde{A}^{ij}
		 + {\textstyle\frac13}\psi^{-4}\tilde\gamma^{ij}K \nonumber\\
	\tilde{A}^{ij} &=& \frac1{2\tilde\alpha}\left(
		(\tilde{\mathbb L}\beta)^{ij} - \tilde{u}^{ij}\right) \\
	\tilde\Delta_{\mathbb L}\beta^i
		- (\tilde{\mathbb L}\beta)^{ij}\tilde\nabla_j\ln\tilde\alpha
		+ {\textstyle\frac43}\tilde\alpha\psi^6\tilde\nabla^iK &=&
		\tilde\alpha\tilde\nabla\left(
			{\textstyle\frac1{\tilde\alpha}}\tilde{u}^{ij}\right)
		+ 16\pi G\tilde\alpha\psi^{10}j^i \nonumber\\
	\tilde\nabla^2\psi - {\textstyle\frac18}\psi\tilde{R}
		- {\textstyle\frac1{12}}\psi^5K^2
		+ {\textstyle\frac18}\psi^{-7}\tilde{A}_{ij}\tilde{A}^{ij}
		&=& -2\pi G\psi^5\rho \nonumber
\end{eqnarray}
In this decomposition (\ref{eq:TS-decomp}), we are free to specify a
symmetric tensor $\tilde\gamma_{ij}$ as the conformal 3-metric, a
symmetric tracefree tensor $\tilde{u}^{ij}$, a scalar function $K$,
\emph{and} the scalar function $\tilde\alpha$.  Solving this set of
equations with appropriate boundary conditions yields initial data
$\gamma_{ij}$ and $K_{ij}$ on a \emph{single} hypersurface.  However,
we also know the following.  \emph{If} we chose to use the shift
vector obtained from solving (\ref{eq:TS-momentum-const}) and the
lapse from (\ref{eq:conf-lapse-def}) via our choice of $\tilde\alpha$
and our solution to the Hamiltonian constraint, then the rate of change
of the physical 3-metric is given by
\begin{eqnarray}\label{eq:TS-gdot-behavior}
	\partial_t\gamma_{ij} &=& u_{ij}
		+ {\textstyle\frac23}\gamma_{ij}\left(
			\bar\nabla_k\beta^k - \alpha K\right) \\
		 &=& \psi^4\left[\tilde{u}_{ij}
		+ {\textstyle\frac23}\tilde\gamma_{ij}\left(
			\tilde\nabla_k\beta^k + 6\beta^k\tilde\nabla_k\ln\psi
			- \psi^6\tilde\alpha K \right)\right]. \nonumber
\end{eqnarray}
This direct information about the consequences of our choices for the
freely specifiable data is something not present in the previous
decompositions.  As we will see later, this framework has been used
to construct initial data that are in quasiequilibrium.

\subsection{Stationary Solutions}
\label{sec:stationary-solutions}

When there is sufficient symmetry present, it is possible to construct
initial data that are in true equilibrium.  These solutions possess at
least two Killing vectors, one that is timelike at large distances and
one that is spatial, representing an azimuthal symmetry.  When these
symmetries are present, solving for the initial data produces a global
solution of Einstein's equations and the solution is said to be
\emph{stationary}.  The familiar Kerr-Neumann solution for rotating
black holes is an example of a stationary solution in vacuum.
Stationary configurations supported by matter are also possible, but
the matter sources must also satisfy the Killing symmetries, in which
case the matter is said to be in hydrostatic
equilibrium~\cite{bardeen-1970}.

The basic approach for finding stationary solutions begins by
simplifying the metric to take into account the symmetries.  Many
different forms have been used for the metric (cf.\ 
Refs.~\cite{bardeen-1970,bardeen-wagoner-1971,bonazzola-schneider-1974,butterworth-ipser-1975,komatsu-etal-1989-I,bonazzola-etal-1993}).
I will use a decomposition that makes comparison with the previous
decompositions straightforward.  First, define the interval as
\begin{equation}\label{eq:stat-interval}
	{\rm d}s^2 = -\psi^{-4}{\rm d}t^2 + \psi^4\left[
		A^2({\rm d}r^2 + r^2{\rm d}\theta^2)
		+ B^2r^2\sin^2\theta({\rm d}\phi + \beta^\phi{\rm d}t)^2
		\right].
\end{equation}
This form of the metric can describe \emph{any} stationary spacetime.
Notice that the lapse is related to the conformal factor by
\begin{equation}\label{eq:stat-lapse}
	\alpha = \psi^{-2},
\end{equation}
and that the shift vector has only one component
\begin{equation}\label{eq:stat-shift}
	\beta^i = (0,0,\beta^\phi).
\end{equation}
I have used the usual conformal decomposition of the 3-metric
(\ref{eq:conformal-metric}) and have written the conformal 3-metric
with two parameters as
\begin{equation}\label{eq:stat-conf-metric}
	\tilde\gamma_{ij} = \left(\begin{array}{ccc}
		A^2	& 0 		& 0	\\
		0	& A^2 r^2	& 0	\\
		0	& 0		& B^2 r^2\sin^2\theta
		\end{array}\right).
\end{equation}
The four functions $\psi$, $\beta^\phi$, $A$, and $B$ are functions of
$r$ and $\theta$ only.

The equations necessary to solve for these four functions are derived
from the constraint equations (\ref{eq:Hamiltonian-const}) and
(\ref{eq:momentum-const}), \emph{and} the evolution equations
(\ref{eq:K-evolution}) and (\ref{eq:g-evolution}).  For the evolution
equations, we use the fact that $\partial_t\gamma_{ij}=0$ and
$\partial_tK_{ij}=0$.  The metric evolution equation
(\ref{eq:g-evolution}) defines the extrinsic curvature in terms of
derivatives of the shift
\begin{equation}\label{eq:stat-excurv-def}
	K_{ij} \equiv \frac1{2\alpha}(
		\bar\nabla_i\beta_j + \bar\nabla_j\beta_i).
\end{equation}
With the given metric and shift, we find that $K=0$ and the divergence
of the shift also vanishes.  This means we can write the tracefree
part of the extrinsic curvature as
\begin{equation}\label{eq:stat-A-def}
	A^{ij} = \psi^{-10}\tilde{A}^{ij} =
		{\textstyle\frac12}\psi^{-2}(\tilde{\mathbb L}\beta)^{ij}.
\end{equation}
We find that the Hamiltonian and momentum constraints take on the
forms given by the thin-sandwich decomposition (\ref{eq:TS-decomp})
with $\tilde{u}^{ij} = K \equiv 0$ and $\tilde\alpha \equiv
\psi^{-8}$.  Only one of the momentum constraint equations is
non-trivial, and we find that the constraints yield elliptic equations
for $\psi$ and $\beta^\phi$.  What remains unspecified as yet are $A$
and $B$ (i.e., the conformal 3-metric).

The conformal 3-metric is determined by the evolution equations for
the traceless part of the extrinsic curvature.  Of these five
equations, one can be written as an elliptic equation for $B$, and two
yield complementary equations that can each be solved by quadrature
for $A$.  The remaining equations are redundant as a result of the
Bianchi identities.

Of course, the clean separation of the equations I have suggested 
above is an illusion.  All four equations must be solved simultaneously,
and clever combinations of the four metric quantities can greatly
simplify the task of solving the system of equations.  This accounts
for the numerous different systems used for solving for stationary
solutions.

\newpage

\section{Black Hole Initial Data}
\label{sec:black-hole-initial}

In this section, we will look at Cauchy initial data that represent
one or more black holes in an asymptotically flat spacetime.  The
majority of these will be either vacuum solutions or solutions of the
Einstein-Maxwell equations.  With no matter to support the
gravitational field, we find that we must usually\footnote{A black
hole can be supported by a compact gravitational
wave~\cite{abrahams-evans-1993}.} use a spacetime with a non-trivial
topology.  It is certainly possible to construct black-hole solutions
supported by
matter~\cite{shapiro-teukolsky-1980,shapiro-teukolsky-1992,arbona-etal-1998},
but it is often desirable to avoid the complications of matter
sources.

This raises a point about solutions of Einstein's equations which we
have not yet mentioned.  When constructing solutions of Einstein's
initial-value equations, we are free to specify the topology of the
initial-data hypersurface.  Einstein's equations of general relativity
place no constraints on the topology of the spacetime they describe or
of spacelike hypersurfaces that foliate it.  For astrophysical black
holes (i.e., black holes in an asymptotically flat spacetime), the
freedom in the choice of the topology has relatively minor
consequences.  The primary effects of different topology choices are
hidden within the black hole's event horizon.

In the sections below, we will explore many of the existing black-hole
solutions and the schemes for generating them.

\subsection{Classic Solutions}
\label{sec:classic-solutions}

\subsubsection{Schwarzschild}
\label{sec:schwarzschild}

Of course, the simplest black-hole solution is the Schwarzschild
solution.  It represents a static spacetime containing a single black
hole that connects two causally disconnected, asymptotically flat
universes.  There are actually many different coordinate
representations of the Schwarzschild solutions.  The simplest
representations are time-symmetric ($K_{ij}=0$), and so exist on a
``maximally embedded'' spacelike hypersurface ($K=0$).  These choices
fix the foliation $\{\Sigma\}$.  Spherical symmetry fixes two of the
three spatial gauge choices.  If we choose an ``areal-radial
coordinate''\footnote{Spheres at constant areal-radial coordinate $r$
have proper area $4\pi r^2$.}, then the interval is written as
\begin{equation}\label{eq:schw-areal-rad-interval}
	{\rm d}s^2 = -\left(1 - {\textstyle\frac{2M}r}\right){\rm d}t^2
		+ \left(1 - {\textstyle\frac{2M}r}\right)^{-1}{\rm d}r^2
		+ r^2({\rm d}\theta^2 + \sin^2\theta{\rm d}\phi^2).
\end{equation}
If we choose an isotropic radial coordinate, then the interval is written
as
\begin{equation}\label{eq:schw-iso-rad-interval}
	{\rm d}s^2 = -\left({\frac{1 - {\textstyle\frac{M}{2\tilde{r}}}}
			{1 + {\textstyle\frac{M}{2\tilde{r}}}}}\right)^2
								{\rm d}t^2
		+ \left(1 + {\textstyle\frac{M}{2\tilde{r}}}\right)^4
			\left({\rm d}\tilde{r}^2
				+ \tilde{r}^2{\rm d}\theta^2
				+ \tilde{r}^2\sin^2\theta{\rm d}\phi^2\right).
\end{equation}
In both (\ref{eq:schw-areal-rad-interval}) and
(\ref{eq:schw-iso-rad-interval}), $M$ represents the mass of the black
hole as measured at spacelike infinity.  Both of these solutions exist
on the same foliation of $t=\mbox{const.}$ slices.  But, notice that
the 3-geometry of the slice associated with
(\ref{eq:schw-iso-rad-interval}) is conformally flat, while the
3-geometry associated with (\ref{eq:schw-areal-rad-interval}) is not.

The solution given in (\ref{eq:schw-iso-rad-interval}) is easily
generated by any of the methods in \S\ref{sec:york-lichn-conf} or
\S\ref{sec:thin-sandw-decomp}.  By choosing a time-symmetric
initial-data hypersurface, we immediately get $K_{ij}=0$, which
eliminates the need to solve the momentum constraints.  If we choose the
conformal 3-geometry to be given by a flat metric (in spherical coordinates in
this case), then the vacuum Hamiltonian constraint
(\ref{eq:conf-Hamiltonian-const-v1}) becomes
\begin{equation}\label{eq:vac-TS-ham}
	\tilde\nabla^2\psi = 0,
\end{equation}
where $\tilde\nabla^2$ is the flat-space Laplace operator.  For the
solution $\psi$ to yield an asymptotically flat physical 3-metric, we
have the boundary condition that $\psi(\tilde{r}\rightarrow\infty) =
1$.  The simplest solution of this equation is
\begin{equation}\label{eq:vac-TS-ham-sol}
	\psi = 1 + \frac{M}{2\tilde{r}},
\end{equation}
where we have chosen the remaining integration constant to give a mass
at infinity of $M$.

We now have full Cauchy initial data representing a single black hole.
If we want to generate a full solution of Einstein's equations, we
must choose a lapse and a shift vector and integrate the evolution
equations~(\ref{eq:K-evolution}) and (\ref{eq:g-evolution}).  In this
case, a reasonable approach for specifying the lapse is to demand that
the time derivative of $K$ vanish.  For the case of $K=0$, this yields
the so-called maximal slicing equation which, for the current
situation, takes the form
\begin{equation}\label{eq:vac-TS-max-slice-eqn}
	\tilde\nabla^2(\alpha\psi) = 0.
\end{equation}
If we choose boundary conditions so that the lapse is frozen on the
event horizon ($\alpha(\tilde{r}=M/2) = 0$) and goes to one at infinity,
we find that the solution is
\begin{equation}\label{eq:vac-TS-iso-lapse}
	\alpha = {\frac{1 - {\textstyle\frac{M}{2\tilde{r}}}}
			{1 + {\textstyle\frac{M}{2\tilde{r}}}}}.
\end{equation}

If we now choose $\beta^i=0$, we find that the left-hand sides of the
evolution equations~(\ref{eq:K-evolution}) and (\ref{eq:g-evolution})
vanish identically, and we have found the static solution of
Einstein's equations given in (\ref{eq:schw-iso-rad-interval}).  We
can, of course, recover the usual Schwarzschild coordinate solution
(\ref{eq:schw-areal-rad-interval}) by using the purely spatial
coordinate transformation $r = \tilde{r}(1 + \frac{M}{2\tilde{r}})^2$.

It is interesting to examine the differences in these two
representations of the Schwarzschild solution.  The isotropic radial
coordinate representation is well behaved everywhere except, it seems,
at $\tilde{r}=0$.  However, even here, the solution is well behaved.
The 3-geometry is invariant under the coordinate transformation
\begin{equation}\label{eq:simple-inversion}
	\tilde{r} \rightarrow \left(\frac{M}2\right)^2\frac1{r^\prime}.
\end{equation}
The event horizon at $\tilde{r}=\frac{M}2$ is a fixed-point set of the
isometry condition (\ref{eq:simple-inversion}) which identifies points
in two causally disconnected, asymptotically flat universes.  We see
that $\tilde{r}=0$ is simply an image of infinity in the other
universe~\cite{brill-lindquist-1963}.

Given our choice for the lapse (\ref{eq:vac-TS-iso-lapse}), which is
frozen on the event horizon, we find that the solution can cover only
the exterior of the black hole.  To cover any of the interior with the
lapse pinned to zero at the horizon would require we use a slice that
is not spacelike everywhere.  This is exactly what happens when the
usual Schwarzschild areal-radial coordinate is used.  At the event
horizon, $r=2M$, there is a coordinate singularity, and inside this
radius the $t=\mbox{const.}$ hypersurface is no longer spacelike.  It
is impossible to perform a Cauchy evolution interior to the event
horizon using the areal-radial coordinate and the given time slicing.

We find that a Cauchy evolution, using the usual Schwarzschild time
slicing that is frozen at the horizon, is capable of evolving only the
region exterior to the black hole's event horizon.  Portions of the
interior of the black hole can be covered by an evolution that begins
with data on a standard Schwarzschild time slice, but the result is
not a time-independent solution.  As we will see later, there are
other slicings of the Schwarzschild spacetime that cover the interior
of the black hole and yield time-independent solutions.

\subsubsection{Time-Symmetric Multi-Hole Solutions}
\label{sec:time-symmetric-multi}

As we saw in \S\ref{sec:schwarzschild}, the simplest approach for
generating initial data is to assume time symmetry and let the
conformal 3-geometry be flat.  We could have generated the exterior
solution for Schwarzschild with the areal-radial coordinate by a
clever choice for the conformal 3-geometry.  The Hamiltonian
constraint is still linear, even if $\tilde\gamma_{ij}$ is not flat.
But there is no obvious motivation for the correct $\tilde\gamma_{ij}$
that would yield the desired solution.

One approach for generating a time-symmetric multi-hole solution is
straightforward.  Brill--Lindquist initial
data~\cite{brill-lindquist-1963,lindquist-1963} again assume a flat
conformal 3-geometry, and the only non-trivial constraint equation is
the Hamiltonian constraint, which again takes the form given in
(\ref{eq:vac-TS-ham}).  But this time, we use the linearity of the
Hamiltonian constraint and choose the solution to be a superposition
of solutions with the form of (\ref{eq:vac-TS-ham-sol}).  More
precisely, we choose the solution
\begin{equation}\label{eq:brill-lindquist-ID}
	\psi = 1 + \sum_{\sigma=1}^N
		\frac{\mu_\sigma}{2|{\bf x} - {\bf C}_\sigma|}.
\end{equation}
Here, $|{\bf x} - {\bf C}_\sigma|$ is a coordinate distance from the
point ${\bf C}_\sigma$ in the Euclidean conformal space, and the
$\mu_\sigma$ are constants related to the masses of the black holes.
Assuming the points ${\bf C}_\sigma$ are sufficiently far apart, this
solution of the initial-value equations represents $N$ black holes
momentarily at rest in ``our'' asymptotically flat universe.  As was
the case for a single black hole, each singular point in the solution,
${\bf x} = {\bf C}_\sigma$, represents infinity in a different,
causally disconnected universe.  In fact, each black hole connects
``our'' universe to a different universe, so that there are $N+1$
asymptotically flat hypersurfaces connected together at the throats of
$N$ black holes.  While we started with a 3-dimensional Euclidean
manifold, the requirement that we delete the singular points, ${\bf
C}_\sigma$, results in a manifold that is not simply connected.  This
solution is often referred to as having a topology with $N+1$
``sheets''.

Brill--Lindquist initial data are very similar to the Schwarzschild
initial data in isotropic coordinates except for one major difference:
the solution does not represent two identical universes that have been
joined together.  The coordinate transformation
(\ref{eq:simple-inversion}) can still be used to show that each pole
in the solution corresponds to infinity on an asymptotically flat
hypersurface.  But, the solution has $N+1$ different asymptotically
flat universes connected together, not two, and ``our'' universe
containing $N$ black holes cannot be isometric to any of the other
universes which contain just one.  Interestingly,
Misner~\cite{misner-1963} found that it is possible to construct a
solution of the vacuum, time-symmetric Hamiltonian constraint
(\ref{eq:vac-TS-ham}) that has two isometric asymptotically flat
hypersurfaces connected by $N$ black holes.  The case of two black
holes that satisfies this isometry condition (often called inversion
symmetry) is usually referred to as ``Misner initial data''.  It has
an analytic representation in terms of an infinite series expansion.
The construction is tedious, and there are several representations of
the solution~\cite{smarr-etal-1976,cook-1991}\footnote{In
Ref.~\cite{cook-1991}, Eq.~(B7), the line ``$1$ for $n=1$'' should
read ``$\alpha^{-1}$ for $n=1$''.}.

Like the Brill--Lindquist data, the non-simply connected topology of
the full manifold is represented on a Euclidean 3-manifold by the
presence of singular points that must be removed.  Brill--Lindquist
data have $N$ singular points, each representing an \emph{image} of
infinity as seen through the throat of the black hole connecting our
universe to that hole's other universe.  Misner's data contain an
infinite number of singular points for each black hole, each
representing an image of one of two asymptotic infinities.  The result
is seen to represent two identical asymptotically flat universes
joined by $N$ black holes.  The two universes join together at the
throats of the $N$ black holes, with each throat being a coordinate
2-sphere in the conformal space.  Data at any point in one universe
are related to data at the corresponding point in the alternate
universe via the same isometry condition (\ref{eq:simple-inversion})
found in the Schwarzschild case.  And, as in the Schwarzschild case,
the 2-sphere throat of each black hole forms a fixed-point set of the
isometry condition.

Both the Brill--Lindquist data and Misner's data represent $N$ black
holes at a moment of time-symmetry (i.e., all the black holes are
momentarily at rest).  Both are conformally flat and the difference in
the topology of the two solutions is hidden from an observer outside
the black holes.  Yet solutions where the holes are chosen to have
the same size and separation yield similar, but physically distinct,
solutions~\cite{cantor-kulkarni-1982,abrahams-price-1996}.

\subsection{General Multi-Hole Solutions}
\label{sec:general-multi-hole}

Time-symmetric black-hole solutions of the constraint equations such
as those described in \S\ref{sec:classic-solutions} are useful as test
cases because they have analytic representations.  However, they have
very little physical relevance.  General time-asymmetric solutions
are needed to represent black holes that are moving and
spinning\footnote{Time-asymmetric solutions are also needed to
represent time-independent solutions that cover the interior of a
black hole.}.  A few approaches for generating general multi-hole
solutions have been explored and, below, we will look at the two
approaches which are direct generalizations of the Misner and
Brill--Lindquist data.  Generalizing these two approaches was the most
natural first step toward constructing general multi-hole solutions.

The first approach to be developed generalized the Misner
approach~\cite{misner-1963}.  It was attractive because an isometry
condition relating the two asymptotically flat universes provides two
useful things.  First, because the two universes are identical,
finding a solution in one universe means that you have the full
solution.  Second, because the throats are fixed-point sets of the
isometry, we can construct boundary conditions on any quantity there.
This allows us to excise the region interior to the spherical throats
from one of the Euclidean background spaces and solve for the initial
data in the remaining volume.

The generalization of the Misner approach seemed preferable to trying
to solve the constraints on $N+1$ Euclidean manifolds stitched
together smoothly at the throats of $N$ black holes.  However, Brandt
and Br{\"u}gmann~\cite{brandt-bruegmann-1997} realized it was possible
to factor out analytically the behavior of the singular points in the
Euclidean manifold of the $N+1$ sheeted approach.  Referred to as the
``puncture'' method, this approach allows us to rewrite the constraint
equations for functions on an $N+1$ sheeted manifold as constraint
equations for new functions on a simple Euclidean manifold.

Another approach tried, which we will not discuss in detail, avoided
the issue of the topology of the initial-data slice entirely.
Developed by Thornburg~\cite{thornburg-1987}, this approach was based
on the idea that only the domain exterior to the apparent horizon of a
black hole is relevant.  The equation describing the location of an
apparent horizon can be rewritten in a form that can be used as a
boundary condition for the conformal factor in the Hamiltonian
constraint equation.  Thus, given a compatible solution to the
momentum constraints, this boundary condition can be used to construct
a solution of the Hamiltonian constraint in the domain exterior to the
apparent horizons of any black holes, with no reference at all to the
topology of the full manifold.

\subsubsection{Bowen--York Data}
\label{sec:bowen-york-data}

The generalization of the Misner approach was developed by York and
his
collaborators~\cite{bowen-york-1980,bowen-1979,bowen-1982,kulkarni-etal-1983,york-1984,kulkarni-1984,bowen-etal-1984,cook-1991}.
The approach is often called the ``conformal-imaging'' method, and the
data are usually referred to as ``Bowen--York'' data.  This approach
begins with a set of simplifying assumptions that is common to all
three of the approaches described above.  These assumptions are that
\begin{eqnarray}\label{eq:Bowen-York-assumptions}
	K &=& 0,\quad\mbox{\it maximal slicing} \nonumber\\
	\tilde\gamma_{ij} &=& f_{ij},\quad\mbox{\it conformal flatness} \\
	\left.\psi\right|_\infty &=& 1.\quad\mbox{\it asymptotic flatness}
		\nonumber
\end{eqnarray}
Here, $f_{ij}$ represents a flat metric in any suitable coordinate
system.  The assumption of conformal flatness means that the
differential operators in the constraints are the familiar flat-space
operators.  More importantly, if we use the conformal
transverse-traceless decomposition (\ref{eq:conf-YL-decomp}), we find
that in vacuum the momentum constraints completely decouple from the
Hamiltonian constraint.

The importance of this last property stems from the fact that York and
Bowen were able to find analytic solutions of this version of the
momentum constraints, solutions that represent a black hole with both
linear momentum and spin~\cite{bowen-york-1980,bowen-1979,bowen-1982}.
If we choose $\tilde{M}^{ij}=0$, then the momentum constraints
(\ref{eq:conf-mom-v2}) become
\begin{equation}\label{eq:Bowen-York-mom-const}
	\tilde\nabla^2V^i + {\textstyle\frac13}\tilde\nabla^i\left(
			\tilde\nabla_jV^j\right) = 0.
\end{equation}
A solution of this equation is
\begin{equation}\label{eq:Bowen-York-simp-vec-sol}
	V^i = -\frac1{4r}\left[7P^i + n^in_jP^j\right]
		+ \frac1{r^2}\epsilon^{ijk}n_jS_k.
\end{equation}
Here, $P^i$ and $S^i$ are vector parameters, $r$ is a coordinate
radius, and $n^i$ is the outward-pointing unit normal of a sphere in
the flat conformal space ($n^i \equiv \frac{x^i}r$).  $\epsilon^{ijk}$
is the 3-dimensional Levi-Civita tensor.

This solution of the momentum constraints yields the tracefree part
of the extrinsic curvature
\begin{eqnarray}\label{eq:Bowen-York-simple-A}
	\tilde{A}_{ij} &=& \frac3{2r^2}\left[P_in_j + P_jn_i
		- (f_{ij} - n_in_j)P^kn_k\right] \\ & &\mbox{}
		+ \frac3{r^3}\left[\epsilon_{ki\ell}S^\ell n^kn_j
			+ \epsilon_{kj\ell}S^\ell n^kn_i\right].\nonumber
\end{eqnarray}
Remarkably, using this solution
(\ref{eq:Bowen-York-simple-A}) and the assumptions in
(\ref{eq:Bowen-York-assumptions}), we can determine the physical
values for the linear and angular momentum of any initial data we can
construct.  The momentum contained in an asymptotically flat initial-data
hypersurface can be calculated from the integral~\cite{york-Essays-GR}
\begin{equation}\label{eq:3+1-momentum-integral}
	\Pi^i\xi^i_{(k)} = \frac1{8\pi}\oint_\infty\left(
		K^j_i - \delta^j_iK\right)\xi^i_{(k)}{\rm d}^2S_j,
\end{equation}
where $\xi^i_{(k)}$ is a Killing vector of the 3-metric
$\gamma_{ij}$\footnote{If $\xi^i_{(k)}$ is a translational Killing
vector, then (\ref{eq:3+1-momentum-integral}) yields the linear
momentum in the direction of that Killing vector.  If $\xi^i_{(k)}$ is
a rotational Killing vector, then (\ref{eq:3+1-momentum-integral})
yields the corresponding angular momentum.}.  Since we are not likely
to have true Killing vectors, we make use of the asymptotic
translational and rotational Killing vectors of the flat conformal
space.  We find from (\ref{eq:3+1-momentum-integral}),
(\ref{eq:Bowen-York-simple-A}), and (\ref{eq:Bowen-York-assumptions})
that the \emph{physical} linear momentum of the initial-data
hypersurface is $P^i$ and the \emph{physical} angular momentum of the
slice is $S^i$.  Furthermore, because the momentum constraints are
linear, we can add any number of solutions of the form of
Eq.~(\ref{eq:Bowen-York-simple-A}) to represent a collection of linear
and angular momentum sources.  The total physical linear momentum of
the initial-data slice will simply be the vector sum of the individual
linear momenta.  The total physical angular momentum cannot be
obtained by simply summing the individual spins because this neglects
the orbital angular momentum of the various sources.  However, the
total angular momentum can still be computed without having to solve
the Hamiltonian constraint~\cite{york-Frontiers}.

The Bowen--York solution for the extrinsic curvature is the starting
point for all the general multi-hole initial-data sets we have
discussed in \S\ref{sec:general-multi-hole}.  However, this solution
is not \emph{inversion symmetric}.  That is, it does not satisfy the
isometry condition that any field must satisfy to exist on a
two-sheeted manifold like that of Misner's solution.  Fortunately,
there is a method of images, similar to that used in electrostatics
but applicable to tensors, that can be used to make any tensor
inversion
symmetric~\cite{misner-1963,bowen-york-1980,kulkarni-etal-1983,kulkarni-1984,york-1984}.

For the conformal extrinsic curvature of a single black hole, there
are two inversion-symmetric solutions~\cite{bowen-york-1980}
\begin{eqnarray}\label{eq:Bowen-York-inversion-sym-A}
	\tilde{A}^\pm_{ij} &=& \frac3{2r^2}\left[P_in_j + P_jn_i
		- (f_{ij} - n_in_j)P^kn_k\right] \nonumber \\ & &\mbox{}
		\mp \frac{3a^2}{2r^4}\left[P_in_j + P_jn_i
		+ (f_{ij} - 5n_in_j)P^kn_k\right] \\ & &\mbox{}
		+ \frac3{r^3}\left[\epsilon_{ki\ell}S^\ell n^kn_j
			+ \epsilon_{kj\ell}S^\ell n^kn_i\right].\nonumber
\end{eqnarray}
Here, $a$ is the radius of the coordinate 2-sphere that is the throat
of the black hole.  Of course, this coordinate 2-sphere is the
fixed-point set of the isometry and is the surface on which we can
impose boundary conditions.  Notice that this radius enters the
solutions only when we make it inversion symmetric.

When the extrinsic curvature represents more than one black hole, the
process for making the solution inversion symmetric is rather complex
and results in an infinite-series solution.  However, in most cases
of interest, the solution converges rapidly and it is straightforward
to evaluate the solution numerically~\cite{cook-1991}.

Given an inversion-symmetric conformal extrinsic curvature, it is
possible to find an inversion-symmetric solution of the Hamiltonian
constraint~\cite{bowen-york-1980}.  Given our assumptions
(\ref{eq:Bowen-York-assumptions}), the Hamiltonian constraint becomes
\begin{equation}\label{eq:Bowen-York-Hamiltonian-const}
	\tilde\nabla^2\psi
		+ {\textstyle\frac18}\psi^{-7}\tilde{A}_{ij}\tilde{A}^{ij}
		= 0.
\end{equation}
The isometry condition imposes a condition on the conformal factor at
the throat of each hole.  This condition takes the
form~\cite{bowen-york-1980}
\begin{equation}\label{eq:Bowen-York-conf-fac-BC}
	\left.n^i_\sigma\tilde\nabla_i\psi\right|_{a_\sigma} =
		- \left.\frac\psi{2r_\sigma}\right|_{a_\sigma},
\end{equation}
where $n^i_\sigma$ is the outward-pointing unit-normal vector to the
$\sigma^{\rm th}$ throat and $a_\sigma$ is the coordinate radius of
that throat.  This condition can be used as a boundary condition when
solving (\ref{eq:Bowen-York-Hamiltonian-const}) in the region exterior
to the throats.

In addition to boundary conditions on the throats, a boundary
condition on the outer boundary of the domain is needed before the
quasilinear elliptic equation in
(\ref{eq:Bowen-York-Hamiltonian-const}) can be solved as a well-posed
boundary-value problem.  This final boundary condition comes from the
fact that we want an asymptotically flat solution.  This implies that
the solution behaves as
\begin{equation}\label{eq:asympt-flat-psi-cond}
	\psi = 1 + \frac{E}{2r} + {\cal O}(r^{-2}),
\end{equation}
where $E$ is the total ADM energy content of the initial-data
hypersurface.  Equation~(\ref{eq:asympt-flat-psi-cond}) can be used
to construct appropriate boundary conditions either at infinity or at
a large, but finite, distance from the black holes~\cite{york-piran-StG}.

\subsubsection{Puncture Data}
\label{sec:puncture-data}

The generalization of the Brill--Lindquist data developed by Brandt
and Br{\"u}g\-mann~\cite{brandt-bruegmann-1997} begins with the same
set of assumptions (\ref{eq:Bowen-York-assumptions}) as the
conformal-imaging approach outlined in \S\ref{sec:bowen-york-data}.
We immediately have (\ref{eq:Bowen-York-simple-A}) from the solution
of the momentum constraints, and we must solve the Hamiltonian
constraint, which again takes the form of
Eq.~(\ref{eq:Bowen-York-Hamiltonian-const}).  At this point, however,
the method of solution differs from the conformal-imaging approach.

Based on the time-symmetric solution, it is reasonable to assume that
the conformal factor will take the form
\begin{equation}\label{eq:puncture-psi-def}
	\psi = \frac1\chi + u, \qquad
		\frac1\chi \equiv \sum_{\sigma=1}^N
			\frac{\mu_\sigma}{2|{\bf x} - {\bf C}_\sigma|}.
\end{equation}
If $u$ is sufficiently smooth, (\ref{eq:puncture-psi-def}) implies
that the manifold will have the topology of $N+1$ asymptotically flat
regions just as in the Brill--Lindquist solution.  In this case,
asymptotic flatness requires that $u = 1 + {\cal O}(r^{-1})$.

Substituting (\ref{eq:puncture-psi-def}) into the Hamiltonian constraint
(\ref{eq:Bowen-York-Hamiltonian-const}) yields
\begin{equation}\label{eq:puncture-Ham-con}
	\tilde\nabla^2u + \eta(1 + \chi u)^{-7} = 0,
\end{equation}
where
\begin{equation}\label{eq:puncture-eta-def}
	\eta = {\textstyle\frac18}\chi^7\tilde{A}_{ij}\tilde{A}^{ij}.
\end{equation}
Near each singular point, or ``puncture'', we find that $\chi \approx
|{\bf x} - {\bf C}_\sigma|$.  From (\ref{eq:Bowen-York-simple-A}), we
see that $\tilde{A}_{ij}\tilde{A}^{ij}$ behaves no worse than $|{\bf
x} - {\bf C}_\sigma|^{-6}$, so $\eta$ vanishes at the punctures at
least as fast as $|{\bf x} - {\bf C}_\sigma|$.

With this behavior, Brandt and
Br{\"u}gmann~\cite{brandt-bruegmann-1997} have shown the existence and
uniqueness of $C^2$ solutions of the modified Hamiltonian constraint
(\ref{eq:puncture-Ham-con}).  The resulting scheme for constructing
multiple black hole initial data is very simple.  The mass and
position of each black hole are parameterized by $\mu_\sigma$ and
${\bf C}_\sigma$, respectively.  Their linear momenta and spin are
parameterized by ${\bf P}_\sigma$ and ${\bf S}_\sigma$ in the
conformal extrinsic curvature (\ref{eq:Bowen-York-simple-A}) used for
each hole.  Finally, the solution for $u$ is found on a simple
Euclidean manifold, with no need for any inner boundaries to avoid
singularities.  This is a great simplification over the
conformal-imaging approach, where proper handling of the inner
boundary is the most difficult aspect of solving the Hamiltonian
constraint numerically~\cite{cook-etal-1993}.

\subsubsection{Problems with These Data}
\label{sec:problems-with-these}

Both the conformal-imaging and puncture methods for generating
multiple black hole initial data allow for completely general
configurations of the relative sizes of the black holes, as well as
their linear and angular momenta.  This does not mean that these
schemes allow for the generation of \emph{all} desired black-hole
initial data.  The two schemes rely on specific assumptions
about the freely specifiable gravitational data.  In particular, they
assume $K=0$, $\tilde{M}^{ij}=0$, and, most importantly, that the
3-geometry is conformally flat.

These choices for the freely specifiable data are not always
commensurate with the desired physical solution.  For example, if we
choose to use either method to construct a single spinning black hole,
we will not obtain the Kerr solution.  The Kerr--Newman solution can
be written in terms of a quasi-isotropic radial coordinate on a $K=0$
time slice~\cite{brandt-seidel-1995-I}.  Let $r$ denote the usual
Boyer--Lindquist radial coordinate and make the standard definitions
\begin{equation}\label{eq:Kerr-functions}
	\rho^2 \equiv r^2 + a^2\cos^2\theta \quad\mbox{and}\quad
	\Delta \equiv r^2 - 2Mr + a^2 + Q^2.
\end{equation}
A quasi-isotropic radial coordinate $\tilde{r}$ can be defined via
\begin{equation}\label{eq:Kerr-isotropic-r-def}
	r = \tilde{r}\left(1 + \frac{M + \sqrt{a^2 + Q^2}}{2\tilde{r}}\right)
		\left(1 + \frac{M - \sqrt{a^2 + Q^2}}{2\tilde{r}}\right).
\end{equation}
The interval then becomes
\begin{equation}\label{eq:Kerr-QI-interval}
	{\rm d}s^2 = -\alpha^2{\rm d}t^2
		+ \psi^4\left[e^{2\mu/3}
			({\rm d}\tilde{r}^2 + \tilde{r}^2{\rm d}\theta^2)
		+ \tilde{r}^2\sin^2\theta e^{-4\mu/3}
			({\rm d}\phi + \beta^\phi{\rm d}t)^2\right],
\end{equation}
with
\begin{eqnarray}\label{eq:Kerr-QI_interval-defs}
	\alpha^2 &=& \frac{\rho^2\Delta}
			{r^2 + a^2)^2 - \Delta a^2\sin^2\theta}
	\nonumber\\
	\beta^\phi &=& -a\frac{r^2 + a^2 - \Delta}
			{(r^2 + a^2)^2 - \Delta a^2\sin^2\theta} \\
	\psi^4 &=& \frac{\rho^{\frac23}
		((r^2 + a^2)^2 - \Delta a^2\sin^2\theta)^{\frac13}}
			{\tilde{r}^2} \nonumber\\
	e^{2\mu} &=& \frac{\rho^4}{(r^2+a^2)^2-\Delta a^2\sin^2\theta}
	\nonumber
\end{eqnarray}

We see immediately that the 3-geometry associated with a
$t=\mbox{const.}$ hypersurface of (\ref{eq:Kerr-QI-interval}) is not
conformally flat.  In fact, Garat and Price~\cite{garat-price-2000}
have shown that in general there is no spatial slicing of the Kerr
spacetime that is axisymmetric, conformally flat, and smoothly goes to
the Schwarzschild solution as the spin parameter $a\rightarrow0$.

Since the Kerr solution is stationary, the inescapable conclusion is
that conformally flat initial data for a single rotating black hole
must also contain some nonvanishing dynamical component.  When we
evolve such data, the system will emit gravitational radiation and
eventually settle down to the Kerr
geometry~\cite{bowen-york-1980,brandt-seidel-1995-II}.  But, it cannot
be the Kerr geometry initially, and it is unlikely that the spurious
gravitational radiation content of the initial data has any desirable
physical properties.  Conformally flat initial data for spinning holes
contain some amount of unphysical ``junk'' radiation\footnote{A
similar conclusion is reached for conformally flat data for a single
black hole with linear momentum~\cite{york-Essays-GR}.}.

The choice of a conformally flat 3-geometry was originally made for
convenience.  Combined with the choice of maximal slicing, these
simplifying assumptions allowed for an analytic solution of the
momentum constraints which vastly simplified the process of
constructing black-hole initial data.  Yet there has been much concern
about the possible adverse physical effects that these choices
(especially the choice of conformal flatness) will have in trying to
study black-hole
spacetimes~\cite{cook-abrahams-1992,lousto-price-1998,rieth-Math-Grav,kley-schaefer-1999,pfeiffer-etal-2000}.
While these conformally flat data sets may still be useful for tests
of black-hole evolution codes, it is becoming widely accepted that the
unphysical initial radiation will significantly contaminate any
gravitational waveforms extracted from evolutions of these data.  In
short, these data are not astrophysically realistic.

The various initial-data decompositions outlined in
\S\ref{sec:york-lichn-conf} and \S\ref{sec:thin-sandw-decomp} are all
capable of producing completely general black-hole initial data sets.
The only limitation of these schemes is our understanding of what
choices to make for the freely specifiable data \emph{and} the
boundary conditions to apply when solving the sets of elliptic
equations.  All these choices will have a critical impact on the
astrophysical significance of the data produced.  It is also important
to remember that similar choices for the freely specifiable data will
result in physically different solutions when applied to the different
schemes.

The first studies of black-hole initial data that are not conformally
flat were carried out by Abrahams {\it et
al.}~\cite{abrahams-etal-1992}.  They looked at the superposition of a
gravitational wave and a black hole.  Using a form of the conformal
metric that allows for so-called Brill waves, they constructed
time-symmetric initial data that were not conformally flat and yet
satisfied the isometry condition (\ref{eq:simple-inversion}) used in
the conformal-imaging method of \S\ref{sec:bowen-york-data}.  These
data were further generalized by Brandt and
Seidel~\cite{brandt-seidel-1996} to include rotating black holes with
a superimposed gravitational wave.  In this case, the data are no
longer time-symmetric yet they satisfy a generalized form of the
isometry condition so that the solution is still represented on two
isometric, asymptotically flat hypersurfaces.

Matzner {\it et al.}~\cite{matzner-etal-1999} have begun to move
beyond conformally flat initial data for binary black holes\footnote{A
related method has been proposed by Bishop {\it et
al.}~\cite{bishop-etal-1998}, but their approach is much different and
outside the current scope of this review.}.  Their proposal is to use
boosted versions of the Kerr metric written in the Kerr--Schild form
to represent each black hole.  Thus, an isolated black hole will have
no spurious radiation content in the initial data.  To construct
solutions with multiple black holes, they propose, essentially, to use
a linear combination of the single-hole solutions.  The resulting
metric can be used as the \emph{conformal} 3-metric, the trace of the
resulting extrinsic curvature can be used for $K$, and the tracefree
part of the resulting extrinsic curvature can be used for
$\tilde{M}^{ij}$.  Their scheme uses York's conformal
transverse-traceless decomposition outlined in
\S\ref{sec:conf-transv-trac}, with the boundary conditions of $\psi=1$
and $V^i=0$ on the horizons of the black holes and conditions
appropriate for asymptotic flatness at large distances from the holes.

The approach outlined by Matzner should certainly yield ``cleaner''
data than the conformally flat data currently available.  For the task
of specifying data for astrophysical black-hole binaries in nearly
circular orbits, it is still true that this new data will not contain
the correct initial gravitational wave content.  Because the black
holes are in orbit, they must be producing a continuous wave-train of
gravitational radiation.  This radiation will not be included in the
method proposed by Matzner {\it et al}.  Also, it is clear that the
boundary conditions being used do not correctly account for the tidal
distortion of each black hole by its companion.  When the black holes
are sufficiently far apart, the radiation from the orbital motion can
be computed using post-Newtonian techniques.  One possibility for
producing astrophysically realistic, binary black-hole initial data is
to use information from these post-Newtonian calculations to obtain
better guesses for $\tilde\gamma_{ij}$, $\tilde{M}^{ij}$, and $K$.

\subsection{Horizon-Penetrating Solutions}
\label{sec:exact-solutions}

We noted in \S\ref{sec:schwarzschild} that the time-independent
maximal slicing of Schwarzschild with isotropic coordinates covers
only the exterior of the black hole.  This is because the time
independence of this gauge requires that the lapse vanish on the
horizon.  It is possible to evolve into the black hole's interior when
starting from initial data constructed in this gauge, but it requires
a choice for the lapse that yields a time-dependent
solution~\cite{smarr-york-1978}.  The time dependence of such a
solution is purely gauge, of course, since the spacetime is static.

It is possible to cover all, or part, of the interior of a single
black hole with a time-independent slicing.  However, doing so seems to
require that we give up the maximal-slicing condition.  To cover the
interior of the black hole, we need a slicing that passes smoothly
through the event horizon.  A convenient way to generate such
solutions is to begin with the metric in standard ingoing-null
coordinates.  If we want to consider a rotating and charged black
hole, then we use the Kerr--Newman geometry in Kerr coordinates:
\begin{eqnarray}\label{eq:ingoing-null-Kerr}
	ds^2 &=& -\left(1 - \frac{2Mr-Q^2}{\rho^2}\right){\rm
        	d}\tilde{V}^2 + 2{\rm d}\tilde{V}{\rm d}r -
        	2\frac{2Mr-Q^2}{\rho^2}a\sin^2\!\theta\,{\rm d}
        	\tilde{V}{\rm d}\tilde\phi \\ & &\mbox{} + \rho^2{\rm
        	d}\theta^2 - 2a\sin^2\!\theta\,{\rm d}r{\rm
        	d}\tilde\phi + \frac{1}{\rho^2}\left[\left(r^2 +
        	a^2\right)^2 - \Delta
        	a^2\sin^2\!\theta\right]\sin^2\!\theta\, {\rm
        	d}\tilde\phi^2, \nonumber
\end{eqnarray}
where $\rho$ and $\Delta$ are defined by Eq.~(\ref{eq:Kerr-functions})
and $\tilde{V}$ is the ingoing-null coordinate.  This metric is
regular at $r = r_\pm\equiv M \pm \sqrt{M^2 - a^2 - Q^2}$, where $r_+$
and $r_-$ are the locations of the event horizon and the Cauchy
horizon, respectively.

This metric can be put into a form suitable for producing
time-independent Cauchy initial data by making coordinate
transformations of the general form
\begin{equation}\label{eq:null-Cauchy-tranforms}
	{\rm d}t = {\rm d}\tilde{V} + f(r){\rm d}r,
		\qquad {\rm d}\phi = {\rm d}\tilde\phi + g(r){\rm d}r,
\end{equation}
where $f$ and $g$ are suitably chosen functions of the radial
coordinate $r$.  There are a few particularly significant solutions
for the general Kerr--Newman geometry, and I will outline these below,
listing the nonzero components of the metric in the standard $3\!+\!1$
format.

\subsubsection{Kerr--Schild Coordinates}
\label{sec:kerr-schild-coord}

A spherical coordinate version of the standard Kerr--Schild coordinate
system is obtained from (\ref{eq:ingoing-null-Kerr}) by using the
coordinate choice (cf.\ Refs.~\cite{marsa-choptuik-1996,cook-scheel-1997})
\begin{equation}\label{eq:KS-coord-def}
	t = \tilde{V} - r \quad\mbox{and}\quad \phi = \tilde\phi.
\end{equation}
The nonzero components of the lapse, shift, and 3-metric are
then given by:
\begin{equation}\label{eq:KS-lapse}
        \alpha^{-2} = 1 + \frac{2Mr-Q^2}{\rho^2},
\end{equation}
\begin{equation}\label{eq:KS-shift-r}
        \beta^r = \alpha^2\frac{2Mr-Q^2}{\rho^2},
\end{equation}
\begin{equation}\label{eq:KS-g_rr}
        \gamma_{rr} = 1 + \frac{2Mr-Q^2}{\rho^2},
\end{equation}
\begin{equation}\label{eq:KS-g_rp}
        \gamma_{r\phi} = -\left[1 + \frac{2Mr-Q^2}{\rho^2}
                \right] a\sin^2\!\theta,
\end{equation}
\begin{equation}\label{eq:KS-g_tt}
        \gamma_{\theta\theta} = \rho^2,
\end{equation}
\begin{equation}\label{eq:KS-g_pp}
        \gamma_{\phi\phi} = \left[r^2 + a^2
                + \frac{2Mr-Q^2}{\rho^2}a^2\sin^2\!\theta\right]\sin^2\!\theta
\end{equation}
Cartesian coordinate components can be obtained from these via the standard
Kerr--Schild coordinate transformations~\cite{MTW}
\begin{equation}\label{eq:KS-Cartesian-trans}
	x + iy = (r + ia)e^{i\phi}\sin\theta \quad\mbox{and}\quad
		z = r\cos\theta.
\end{equation}
This yields the implicit definition of $r$ from
\begin{equation}\label{eq:KS-implicit-r}
	r^4 - r^2(x^2 + y^2 + z^2 - a^2) - a^2z^2 = 0,
\end{equation}
with $r > 0$ and $r=0$ on the disk described by $z=0$ and $x^2 + y^2
\le a^2$.

\subsubsection{Harmonic Coordinates}
\label{sec:harmonic-coordinates}

Harmonic time slicing is integral to some hyperbolic formulations of
general relativity, and a time-independent harmonic slicing of the
Kerr--Newman geometry does
exist~\cite{bona-masso-1988,cook-scheel-1997}.  The harmonic time
slicing condition is $\Box{t} = 0$, which can be written
\begin{equation}\label{eq:harm-time-cond}
	\frac1{\sqrt{-g}}\partial_\mu(\sqrt{-g}g^{0\mu}) = 0.
\end{equation}
This equation is satisfied by using the coordinate choice
\begin{equation}\label{eq:Harm-coord-def}
	t = \tilde{V} - r + 2M\ln\left|\frac{2M}{r - r_-}\right|,
		 \qquad \phi = \tilde\phi.
\end{equation}
The nonzero components of the lapse, shift, and 3-metric are then
given by:
\begin{equation}\label{eq:Harm-lapse}
        \alpha^{-2} = 1 + \frac{2Mr-Q^2}{\rho^2}
                                \left(\frac{r+r_+}{r - r_-}\right)
                + \frac{r_+^2 + a^2}{\rho^2}\left(\frac{2M}{r - r_-}\right)
\end{equation}
\begin{equation}\label{eq:Harm-shift-r}
        \beta^r = \alpha^2\frac{r_+^2 + a^2}{\rho^2}
\end{equation}
\begin{equation}\label{eq:Harm-shift-p}
        \beta^\phi  = - \alpha^2 \frac{a}{\rho^2}
                                \left(\frac{2M}{r - r_-}\right)
\end{equation}
\begin{equation}\label{eq:Harm-g_rr}
        \gamma_{rr} = \left[2 - \left(1 - \frac{2Mr-Q^2}{\rho^2}\right)
                \frac{r+r_+}{r - r_-}\right]
                \frac{r+r_+}{r - r_-}
\end{equation}
\begin{equation}\label{eq:Harm-g_rp}
        \gamma_{r\phi} = -\left[1 + \frac{2Mr-Q^2}{\rho^2}
                \left(\frac{r+r_+}{r - r_-}\right)\right] a\sin^2\!\theta
\end{equation}
\begin{equation}\label{eq:Harm-g_tt}
        \gamma_{\theta\theta} = \rho^2
\end{equation}
\begin{equation}\label{eq:Harm-g_pp}
        \gamma_{\phi\phi} = \left[r^2 + a^2
                + \frac{2Mr-Q^2}{\rho^2}a^2\sin^2\!\theta\right]\sin^2\!\theta
\end{equation}
Cartesian coordinate components can be obtained from these via the
standard Kerr--Schild coordinate transformations
(\ref{eq:KS-Cartesian-trans}) and (\ref{eq:KS-implicit-r}).  However,
for the harmonic slicing, the $t = \mbox{const.}$ hypersurface is
spacelike only outside the Cauchy horizon at $r>r_-$.

Fully harmonic coordinates ($\Box{x^\mu}=0$) can be defined when
Cartesian spatial coordinates are used by employing a variation of the
standard Kerr--Schild coordinate transformations~\cite{cook-scheel-1997}
\begin{equation}\label{eq:full-harm-Cartesian-trans}
	x + iy = (r - m + ia)e^{i\phi}\sin\theta \quad\mbox{and}\quad
		z = (r - m)\cos\theta.
\end{equation}
This yields the implicit definition of $r$ from
\begin{equation}\label{eq:full-harm-implicit-r}
	(r-m)^4 - (r-m)^2(x^2 + y^2 + z^2 - a^2) - a^2z^2 = 0.
\end{equation}
Fully harmonic coordinates are useful because applying a boost to a
harmonically sliced black hole yields a solution that satisfies
(\ref{eq:harm-time-cond}) only if the black hole is written in fully
harmonic coordinates.  In this case, the boosted solution also
satisfies the fully harmonic coordinate conditions.

\subsubsection{Generalized Painlev{\'e}--Gullstrand Coordinates}
\label{sec:painl-gullstr-coord}

The Painlev{\'e}--Gullstrand gauge choice for the Schwarzschild
geometry has been rediscovered many times because of its simple form
(cf.\ Refs.~\cite{painleve-1921,gullstrand-1922,robertson-noonan-1968,kraus-wilczek-1994,lake-1994}).
It is another time-independent solution, but the 3-geometry is
completely flat (not simply conformally flat).  The lapse is one in
this gauge, and all of the information regarding the curvature of
spacetime is contained in the shift.  The Painlev{\'e}--Gullstrand
gauge also has an intuitive physical
interpretation~\cite{martel-poisson-2000}.  An observer starting at
rest at infinity and freely falling will trace out a world line that
is everywhere orthogonal to the $t=\mbox{const.}$ hypersurfaces in
Painlev{\'e}--Gullstrand coordinates.

A generalization of the Painlev{\'e}--Gullstrand gauge derived by
Doran~\cite{doran-2000} includes the Kerr spacetime, and the extension
of this solution to the full Kerr--Newman spacetime is trivial.  In the
limit that $a$ and $Q$ vanish, this solution reduces to the
Painlev{\'e}--Gullstrand gauge.  The coordinate transformation is
written most easily as
\begin{equation}\label{eq:PG-coord-def}
	{\rm d}t = {\rm d}\tilde{V}
		- \frac{{\rm d}r}{1 + \sqrt{\frac{2Mr-Q^2}{r^2 + a^2}}},
	\qquad {\rm d}\phi = {\rm d}\tilde\phi
		- \frac{a}{r^2 + a^2}
			\frac{{\rm d}r}{1 + \sqrt{\frac{2Mr-Q^2}{r^2 + a^2}}}
\end{equation}
The nonzero components of the lapse, shift, and 3-metric are then
given by:
\begin{equation}\label{eq:PG-lapse}
        \alpha^{-2} = 1
\end{equation}
\begin{equation}\label{eq:PG-shift-r}
        \beta^r = \alpha^2\sqrt{\frac{r^2 + a^2}{\rho^2}}
				\sqrt{\frac{2Mr-Q^2}{\rho^2}}
\end{equation}
\begin{equation}\label{eq:PG-g_rr}
        \gamma_{rr} = \frac{\rho^2}{r^2 + a^2}
\end{equation}
\begin{equation}\label{eq:PG-g_rp}
        \gamma_{r\phi} = -\left[\sqrt{\frac{\rho^2}{r^2 + a^2}}
				\sqrt{\frac{2Mr-Q^2}{\rho^2}}\right]
			a\sin^2\!\theta
\end{equation}
\begin{equation}\label{eq:PG-g_tt}
        \gamma_{\theta\theta} = \rho^2
\end{equation}
\begin{equation}\label{eq:PG-g_pp}
        \gamma_{\phi\phi} = \left[r^2 + a^2
                + \frac{2Mr-Q^2}{\rho^2}a^2\sin^2\!\theta\right]\sin^2\!\theta
\end{equation}
Notice that the lapse remains one, but the 3-geometry is no longer
flat when the black hole is spinning.  Cartesian coordinate components
can be obtained from these via the standard Kerr--Schild coordinate
transformations (\ref{eq:KS-Cartesian-trans}) and
(\ref{eq:KS-implicit-r}).  Like the Kerr--Schild time slicing, a
$t=\mbox{const.}$ slice of the generalized Painlev{\'e}--Gullstrand
gauge remains spacelike for all $r\ge0$.

\subsection{Quasicircular Binary Data}
\label{sec:quasicirc-binary}

One of the primary driving forces behind the development of black-hole
initial data has been the two-body problem of general relativity: the
inspiral and coalescence of a pair of black holes.  This problem is of
fundamental importance.  Not only is the relativistic two-body problem
the most fundamental dynamical problem of general relativity, it is
also considered one of the most likely candidates for observation with
the upcoming generation of gravitational wave laser interferometers.
Because of the circularizing effects of gravitational radiation
damping, we expect the orbits of most tight binary systems to have
small eccentricities.  It is therefore desirable to have a method that
can discern which data sets, within the very large parameter space of
binary black-hole initial-data sets, correspond to black-hole binaries
in a nearly circular (quasicircular) orbit.

Currently, only one approach has been developed for locating
quasicircular orbits in a parameter space of binary black-hole
initial data~\cite{cook-1994}.  It is based on the fact that minimizing
the energy of a binary system while keeping the orbital angular
momentum fixed will yield a circular orbit in Newtonian gravity.  The
idea does not hold strictly for general relativistic binaries since
they emit gravitational radiation and cannot be in equilibrium.
However, for orbits outside the innermost stable circular orbit, the
gravitational radiation reaction time scale is much longer than the
orbital period. Thus it is a good approximation to treat such binaries
as an equilibrium system.  Called an ``effective potential method'',
this approach was used originally to find the quasicircular orbits
and innermost stable circular orbit (ISCO) for equal-sized nonspinning
black holes~\cite{cook-1994}.  In this work, the initial data for
binary black holes were computed using the conformal-imaging approach
outlined in \S\ref{sec:bowen-york-data}.  The approach was also
applied to binary black-hole data computed using the puncture
method~\cite{baumgarte-2000}, where similar results were found.
Configurations containing a pair of equal-sized black holes with spin
also have been examined~\cite{pfeiffer-etal-2000}.  In this case, the
spins of the black holes are equal in magnitude, but are aligned
either parallel to, or anti-parallel to, the direction of the orbital
angular momentum.

The approach defines an ``effective potential'' based on the binding
energy of the binary.  The binding energy is defined as
\begin{equation}\label{eq:binding-energy}
	E_b \equiv E_{\rm ADM} - M_1 - M_2,
\end{equation}
where $E_{\rm ADM}$ is the total ADM energy of the system measured at
infinity, and $M_1$ and $M_2$ are the masses of the individual black
holes.  Quasicircular orbital configurations are obtained by
minimizing the effective potential (defined as the nondimensional
binding energy $E_b/\mu$ (where $\mu \equiv M_1M_2/(M_1 + M_2)$) as a
function of separation, while keeping the the ratio of the masses of
the black holes $M_1/M_2$, the spins of the black holes ${\bf
S}_1/M_1^2$ and ${\bf S}_2/M_2^2$, and the total angular momentum
$J/M_1M_2$ constant.

This approach is limited primarily by the ambiguity in defining the
individual masses of the black holes, $M_1$ and $M_2$.  There is no
rigorous definition for the mass of an individual hole in a binary
configuration and some approximation must be made here.  There is also
no rigorous definition for the individual spins of holes in a binary
configuration.  The problem of defining the individual masses becomes
particularly pronounced when the holes are very close together (see
Ref.~\cite{pfeiffer-etal-2000}).  The limiting choice of conformal
flatness for the 3-geometry also has proven to be problematic.  The
effects of this choice have been clearly seen in the case of
quasicircular orbits of spinning black
holes~\cite{pfeiffer-etal-2000}, but it is also believed to be a
serious problem for any binary configuration because binary
configurations are not conformally flat at the second post-Newtonian
order~\cite{rieth-Math-Grav}.

To date, the results of the effective-potential method have not
matched well to the best result from post-Newtonian
approximations~\cite{Damour-etal-2000}\footnote{We note that for the
preferred choice of $\omega_{\rm static}\approx-9$ the gauge invariant
parameters of the ISCO found in Ref.~\cite{Damour-etal-2000} do not
match well with the same parameter determined via the
effective-potential method in Ref.~\cite{cook-1994}.  However, if $0 <
\omega_{\rm static}<10$, the agreement is much better.}.  It will be
interesting to see if the results from post-Newtonian approximations
and numerical initial-data sets converge, especially when the
approximation of conformal flatness is eliminated.

\newpage

\section{Neutron-Star Initial Data}\label{sec:neutron-star-initial}

The construction of initial data for neutron stars requires that the
state of the neutron-star matter be specified before the gravitational
data can be determined.  Of course, a solution of the gravitational
constraint equations can be found in principal for any given energy
density $\rho$ and momentum density $j^i$.  But, with neutron-star
solutions, we are usually interested in situations where the matter
is in (or nearly in) hydrostatic equilibrium and the gravitational
fields are also in (or nearly in) equilibrium.

\subsection{Hydrostatic Equilibrium}\label{sec:hydr-equil}

For a neutron star to be in true equilibrium, the spacetime must be
stationary as discussed in \S\ref{sec:stationary-solutions}.  This
means that the spacetime possesses both ``temporal'' and ``angular''
Killing vectors (cf.\ Ref.~\cite{butterworth-ipser-1976}).  If the
matter is also to be in equilibrium, then the 4-velocity of the matter
$u^\mu$ must be a linear combination of these two Killing vectors.  If
we use coordinates as defined in (\ref{eq:stat-interval}) with the
angular Killing vector in the $\phi$ direction, then
\begin{equation}\label{eq:equilib-4-velocity}
	u^\mu = u^t[1,0,0,\Omega].
\end{equation}
Here, $u^t$ and $\Omega$ are functions of $r$ and $\theta$ only.
$\Omega$ is the angular velocity of the matter as measured at
infinity.

It is common to define $v$ as the relative velocity between the matter
and a normal observer (often called a zero angular momentum observer)
so that
\begin{equation}\label{eq:v-ZAMO-def}
	 \frac1{\sqrt{1 - v^2}} = -n_\mu u^\mu = \alpha u^t.
\end{equation}
The velocity $v$ is then fixed by the normalization condition $u^\mu
u_\mu = -1$.

If we assume that the matter source is a perfect fluid, then the
stress-energy tensor is given by
\begin{equation}\label{eq:perf-fluid-T}
	T^{\mu\nu} = (\varepsilon + P)u^\mu u^\nu + Pg^{\mu\nu},
\end{equation}
where $\varepsilon$ and $P$ are the total energy density and pressure,
respectively, as measured in the rest frame of the fluid.  The
vanishing of the divergence of the stress-energy tensor yields the
equation of hydrostatic equilibrium (often referred to as the 
relativistic Bernoulli equation).  In differential form, this is
\begin{equation}\label{eq:hydro-equilib}
	{\rm d}P - (\varepsilon + P)
		({\rm d}\ln u^t - u^t u_\phi{\rm d}\Omega) = 0.
\end{equation}
If the fluid is barytropic\footnote{For a barytropic fluid, the
entropy per baryon and the fractional abundances of the different
nuclear species are determined uniquely by the distribution of
baryons. In this case, the total energy density $\varepsilon$ can be
expressed as a function of the pressure $P$ (cf.\ 
Ref.~\cite{butterworth-ipser-1976}).}, then we can define the
relativistic \emph{enthalpy} as
\begin{equation}\label{eq:enthalpy}
	h(P) \equiv \exp\left[\int_0^P\frac{dP}{\varepsilon + P}\right],
\end{equation}
and rewrite the relativistic Bernoulli equation as
\begin{equation}\label{eq:barytropic-bernoulli}
	\ln h(P) = \ln h_0 + \ln u^t - \ln u^t_0
		- \int_{\Omega_0}^\Omega u^tu_\phi d\Omega.
\end{equation}
The constants $h_0$, $u^t_0$, and $\Omega_0$ are the values their
respective quantities have at some reference point, often taken to
be the surface of the neutron star at the axis of rotation.  When
uniform rotation is assumed (${\rm d}\Omega=0$),
Eq.~(\ref{eq:barytropic-bernoulli}) is rather easy to solve.  The
case of differential rotation is somewhat more complicated.  An
integrability condition of (\ref{eq:hydro-equilib}) requires that
$u^tu_\phi$ be expressible as a function of $\Omega$, so
\begin{equation}\label{eq:rotation-law}
	u^tu_\phi \equiv F(\Omega).
\end{equation}
$F(\Omega)$ is a specifiable function of $\Omega$ which determines the
rotation law that the neutron star must
obey~\cite{butterworth-ipser-1976}.

\subsection{Isolated Neutron Stars}
\label{sec:isol-neutr-stars}

The simplest models of isolated neutron stars are static (i.e.,
nonrotating), spherically symmetric models that can be constructed,
given a suitable equation of state, by solving the
Oppenheimer--Volkoff (OV) equations~\cite{oppenheimer-volkoff-1939}:
\begin{eqnarray}\label{eq:OV-areal}
	\frac{dP}{dr} &=& -\frac{(\varepsilon + P)(m + 4\pi r^3P)}
					{r(r - 2m)}  \nonumber \\
	\frac{dm}{dr} &=& 4\pi r^2\varepsilon \\
	\frac{d\alpha}{dr} &=&
		\frac{\alpha(m + 4\pi r^3\varepsilon)}{r(r - 2m)} \nonumber
\end{eqnarray}
for $0\le r\le R$, where $R$ is the radius of the surface of the star.
Here, $r$ is an areal radius and $m(r)$ is the mass inside radius $r$.
Exterior to the surface of the star, the metric is the standard
Schwarzschild metric as in Eq.~(\ref{eq:schw-areal-rad-interval}) with
$M \equiv m(R)$.  Interior to the surface of the star, the metric is
\begin{equation}\label{eq:OV-interior-metric}
	{\rm d}s^2 = - \alpha^2{\rm d}t^2
		+ \left({1 - \textstyle\frac{2m(r)}{r}}\right)^{-1}{\rm d}r^2
		+ r^2({\rm d}\theta^2 + \sin^2\theta{\rm d}\phi^2).
\end{equation}
The boundary conditions are that
$m(0)=0$, $\varepsilon(0)$ is some chosen constant $\varepsilon_0$,
and $\alpha(R) = \sqrt{1 - \frac{2M}R}$.  The solutions of this
equation form a one-parameter family, parameterized by $\varepsilon_0$
which determines how relativistic the system is.  A method for solving
these equations in both the areal coordinate $r$ and an isotropic
radial coordinate $\tilde{r}$ can be found in
Ref.~\cite{butterworth-1976}.

More generally, isolated neutron stars will be rotating.  If the
neutron stars are uniformly rotating, then, for any given
equation of state, the solutions form a two-parameter family.  These
models can be parameterized by their central density, which determines
how relativistic they are, and by the amount of rotation.  If the
models are allowed to have differential rotation, then some rotation
law must be chosen.

To construct a neutron-star model, the equations for a stationary
solution of Einstein's equations outlined in
\S\ref{sec:stationary-solutions} must be solved self-consistently with
the equations for hydrostatic equilibrium of the matter outlined above
in \S\ref{sec:hydr-equil}.  The equations that must be solved depend
on the form of the metric chosen, and numerous formalisms and
numerical schemes have been used.  An incomplete list of references to
work on constructing neutron-star models include
\cite{wilson-1972,bonazzola-schneider-1974,butterworth-ipser-1975,butterworth-1976,butterworth-ipser-1976,butterworth-1979,Friedman-etal-1986,komatsu-etal-1989-I,komatsu-etal-1989-II,cook-etal-1992,cook-etal-1994-P,cook-etal-1994-R,stergioulas-friedman-1995,bonazzola-etal-1993,gourgoulhon-bonazzola-1993,bocquet-etal-1995,bonazzola-etal-1998}.
Further review information on neutron-star models can be found in
Refs.~\cite{stergioulas-1998,friedman-ipser-1992,friedman-ipser-1992-E}.

\subsection{Neutron-Star Binaries}
\label{sec:neutr-star-binar}

Neutron-star binaries, like any relativistic binary system, cannot
exist in a true equilibrium configuration since they must emit
gravitational radiation.  But, as is true for black-hole binary data,
for orbits outside the innermost stable circular orbit, the
gravitational radiation reaction time scale is much longer than the
orbital period and it is a reasonable approximation to consider the
stars to be in a quasiequilibrium state.

A binary configuration obviously lacks the azimuthal symmetry that was
assumed in the discussions of stationary solutions of Einstein's
equations in \S\ref{sec:stationary-solutions} and hydrostatic
equilibrium in \S\ref{sec:hydr-equil}.  Fortunately, the condition of
hydrostatic equilibrium requires only the presence of a single,
timelike Killing vector.  With the assumption that gravitational
radiation is negligible, we can assume that the matter is in some
equilibrium state as viewed from the reference frame that is rotating
along with the binary.  That is, if the binary has a constant orbital
angular velocity of $\Omega$, then the time vector in the rotating
frame $t^{\prime\mu}$ is a Killing vector\footnote{Bonazzola {\it et
al}.~\cite{bonazzola-etal-1997} refer to this as a \emph{helicoidal}
Killing vector.} and it is related to the time vector in the rest
frame of the binary $t^\mu$ by
\begin{equation}\label{eq:co-rotating-killing-vec}
	t^{\prime\mu} = t^\mu + \Omega \xi^\mu,
\end{equation}
where $\xi^\mu$ is a generator of rotations about the rotation
axis\footnote{In spherical coordinates, $\vec\xi =
\partial/\partial\phi$.  In Cartesian coordinates, rotation about the
$z$ axis would be represented by $\xi^\mu = (0,-y,x,0)$.}.

Two equilibrium states for the matter have been explored in the
literature.  The simplest case is that of \emph{co}-rotation, where
the 4-velocity of the matter is proportional to $t^{\prime\mu}$.  In
this case, the matter is at rest in the frame of reference rotating
with the binary system, the \emph{corotating reference frame}.  The
second equilibrium state is that of \emph{counter}-rotation, where
there is no rotation in the \emph{rest frame of the binary}.  We will
explore these two cases further below.

Stationarity of the gravitational field, unlike hydrostatic
equilibrium, requires the presence of separate timelike and azimuthal
Killing vectors.  For the case of a binary system, there is no unique
definition of quasiequilibrium.  The earliest work on constructing
quasiequilibrium solutions of Einstein's equations stems from work by
Wilson and
Mathews~\cite{wilson-mathews-Frontiers,wilson-mathews-1995,wilson-etal-1996},
and others have explored similar
schemes~\cite{baumgarte-etal-1997,bonazzola-etal-1997,baumgarte-etal-1998-model}.
Although written in slightly different forms, the system of equations
for the gravitational fields in all of these schemes are fundamentally
identical.  While they were developed before the thin-sandwich
decomposition of \S\ref{sec:thin-sandw-decomp}, the thin-sandwich
decomposition (see Eq.~(\ref{eq:TS-decomp})) offers the easiest way to
interpret this approach.  We consider ourselves to be in the
corotating reference frame so that our time vector is $t^{\prime\mu}$.
To make the transition back to the rest frame of the binary as easy as
possible, we write the shift vector of our $\!3+\!1$ decomposition as
\begin{equation}\label{eq:WM-shift}
	B^i = \beta^i + \Omega\xi^i,
\end{equation}
so that $\beta^i$ remains as the shift vector of the $3\!+\!1$
decomposition made with respect to the rest frame of the binary
system.

The primary assumptions are that the conformal 3-metric
$\tilde\gamma_{ij}$ is flat, the initial-data slice is maximal so that
$K=0$, and $\tilde{u}^{ij}=0$.  We see from (\ref{eq:TS-conf-u-def})
that the last assumption implies that the conformal 3-geometry is
\emph{instantaneously} stationary as seen in the corotating reference
frame.  The final choice that must be made is for the conformally
rescaled lapse $\tilde\alpha$.  An elliptic equation for the lapse can
be obtained by demanding the trace of the extrinsic curvature $K$ also
be instantaneously stationary in the corotating reference frame.  This
is the so-called \emph{maximal slicing} condition on the lapse.  For
the particular assumptions we have made here, this equation can be
written
\begin{equation}\label{eq:WM-maximal-slicing}
	\tilde\nabla^2(\tilde\alpha\psi^7) =
		(\tilde\alpha\psi^7)\left[
		\frac78\psi^{-7}\tilde{A}_{ij}\tilde{A}^{ij}
			+ 2\pi G\psi^4(\rho + 2S)\right].
\end{equation}

It is interesting to note that, for a conformally flat 3-geometry,
\begin{equation}\label{eq:CF-axial-simp}
	(\tilde{\mathbb L}B)^{ij} = (\tilde{\mathbb L}\beta)^{ij},
\end{equation}
so $\Omega$ does not appear in the equations for the gravitational
fields except in the matter terms and possibly in boundary conditions.
Equations~(\ref{eq:TS-decomp}) and (\ref{eq:WM-maximal-slicing}) can
be solved for the gravitational fields on an initial-data
hypersurface, given values for the matter terms and appropriate
boundary conditions.

\emph{If} we choose the matter so that it is in hydrostatic
equilibrium with respect to the pseudo-Killing vector $t^{\prime\mu}$,
then these equations for the gravitational fields will yield data that
are in quasiequilibrium in the sense that $\tilde{\gamma}_{ij}$ and
$K$ are both \emph{instantaneously} stationary.

For corotating binaries, the matter is at rest in the corotating
reference frame of the binary.  It is \emph{rigidly} rotating and
hydrostatic equilibrium is specified by solving the relativistic
Bernoulli equation (\ref{eq:barytropic-bernoulli}), with ${\rm
d}\Omega=0$, self-consistently with the equations for the
gravitational fields.

For counterrotating binaries, the matter is not rotating in the rest
frame of the binary.  Counterrotating equilibrium configurations can
be obtained by assuming the matter to have \emph{irrotational
flow}~\cite{teukolsky-1998,shibata-1998}.  As long as the flow is
\emph{isentropic}, we can express the enthalpy (\ref{eq:enthalpy})
as\footnote{For isentropic flow, the thermodynamic identity reduces to
${\rm d}h = \frac1{\rho_0}{\rm d}P$.}
\begin{equation}\label{eq:isentropic-enthalpy}
	h = \frac{\varepsilon + P}{\rho_0},
\end{equation}
where $\rho_0$ is the rest-mass density.  For irrotational flow, the
vorticity of the fluid (cf.\ Ref.~\cite{teukolsky-1998}) is zero.
Combining this with the Euler equation, we find that the 4-velocity
of the fluid can be expressed as
\begin{equation}\label{eq:irrotational-flow-def}
	hu_\mu = \nabla_\mu\varphi,
\end{equation}
where $\varphi$ is the velocity potential, or flow field.
Equation~(\ref{eq:irrotational-flow-def}), together with the
normalization condition $u^\mu u_\mu=-1$, automatically satisfies the
Euler equation and we are left with the continuity equation which must
be satisfied,
\begin{equation}\label{eq:continuity}
	\nabla_\mu(\rho_0u^\mu) = 0.
\end{equation}
The continuity equation~(\ref{eq:continuity}), the normalization
condition, and Eq.~(\ref{eq:irrotational-flow-def}) yield
\begin{equation}\label{eq:irrot-flow-eqns}
	\nabla^2\varphi = -(\nabla^\mu\varphi)\nabla_\mu\ln(\rho_0/h)
		\quad\mbox{with}\quad
	h = \sqrt{-(\nabla^\nu\varphi)(\nabla_\nu\varphi)}.
\end{equation}

Stationarity (or quasistationarity) in the corotating reference frame
requires that
\begin{equation}\label{eq:irrot-stationary-cond}
	hu_\mu t^{\prime\mu} = -C,
\end{equation}
where $C$ is a positive constant.  Now, in terms of the $3\!+\!1$
decomposition with $B^i$ as the shift vector (see
Eq.~(\ref{eq:WM-shift})), we find that the Bernoulli equation is written
\begin{equation}\label{eq:irrot-bernoulli}
	h^2 = -(\bar\nabla^i\varphi)\bar\nabla_i\varphi
		+ {\textstyle\frac1{\alpha^2}}\left(C
				+ B^j\bar\nabla_j\varphi\right)^2.
\end{equation}
The flow field $\varphi$ must satisfy
\begin{eqnarray}\label{eq:irrot-flow-eqn}
	\bar\nabla^2\varphi
		- B^i\bar\nabla_i\left(\frac{\lambda}{\alpha^2}\right)
		- \frac{\lambda}{\alpha}K &=& 
		-\left(\bar\nabla^i\varphi - \frac{\lambda}{\alpha^2}B^i\right)
			\bar\nabla_i\ln\left(\frac{\alpha\rho_0}{h}\right) \\
	\lambda &\equiv& C + B^i\bar\nabla_i\varphi \nonumber
\end{eqnarray}
subject to the boundary condition\footnote{This boundary condition
comes from the fact that the fluid motion at the surface of the star
must be tangent to the surface, $u^\mu\nabla_\mu\rho_0 =
0$.} at the surface of the flow
\begin{equation}\label{eq:irrot-flow-BC}
	\left.\left(\bar\nabla^i\varphi - \frac{\lambda}{\alpha^2}B^i\right)
			\bar\nabla_i\rho_0\right|_{\rm surf} = 0.
\end{equation}
Solving Eqs.~(\ref{eq:irrot-bernoulli}) and (\ref{eq:irrot-flow-eqn})
self-consistently with the equations for the gravitational fields
yields a counterrotating binary in hydrostatic equilibrium.

As mentioned above, the earliest work on neutron-star binaries was
carried out by Wilson and
Mathews~\cite{wilson-mathews-Frontiers,wilson-mathews-1995}.  Wilson
{\it et al}.~\cite{wilson-etal-1996} describe their approach for
generating initial data for equilibrium neutron-star binaries.  In
these early works, the equation of hydrostatic equilibrium was not
used.  Rather, an initial guess for the density profile was chosen and
the full hydrodynamic system was evolved with viscous damping until
equilibrium was reached.  During each step of the hydrodynamic
evolution, the equations for the gravitational fields were resolved.
The resulting data represented neither strictly co- nor
counter-rotating binary neutron stars.  This work led to the
controversial result~\cite{mathews-etal-1998} that each neutron star in
the binary may become radially unstable and collapse prior to the
merger of the pair of stars.  While an error was found in this
work~\cite{flanagan-1999,mathews-wilson-2000} with the result that the
signature of collapse is significantly weaker, the controversy has not
yet been completely resolved.

The first use of corotating hydrostatic equilibrium with the
Wilson--Mathews approach for specifying the gravitational fields was
by Cook {\it et al}.~\cite{cook-etal-1996} for the test case of an
isolated neutron star.  This approach was then used to study
corotating neutron-star binaries by Baumgarte {\it et
al}.~\cite{baumgarte-etal-1997,baumgarte-etal-1998-model} and by
Marronetti {\it et al}.~\cite{marronetti-etal-1998}.
Interestingly, turning-point methods for detecting \emph{secular}
instabilities~\cite{sorkin-1981,sorkin-1982,friedman-etal-1988} can be
applied to the case of corotating
binaries~\cite{baumgarte-etal-1998-stab}.

Corotating binary configurations are relatively easy to construct.
However, it is believed that the viscosity of neutron-star matter is
not large enough to allow for synchronization of the spin with the
orbit~\cite{kochanek-1992,bildsten-cutler-1992}.  But, if the initial
spins of the neutron stars are not too large, close binaries should be
well approximated by irrotational models.  Bonazzola {\it et
al}.~\cite{bonazzola-etal-1997} (as corrected by
Asada~\cite{asada-1998}) developed the first approach for constructing
counterrotating binary configurations.  However, simpler formulations
of irrotational flow were developed independently by
Teukolsky~\cite{teukolsky-1998} and Shibata~\cite{shibata-1998}, and
Gourgoulhon~\cite{gourgoulhon-1998} showed that all three approaches
were equivalent.  Numerical solutions of the equations for
irrotational flow coupled to the equations for the gravitational
fields are more difficult to construct than those for corotation
because of the boundary condition (\ref{eq:irrot-flow-BC}) on the flow
field that must be applied on the surface of each neutron star.  This
boundary condition is particularly difficult to implement because the
location of the surface of the star is not known {\it a priori}, and
will move as the equations are being solved.  The first models of
irrotational binary neutron stars were constructed by Bonazzola {\it
et al}.~\cite{bonazzola-etal-1999}, Marronetti {\it et
al}.~\cite{marronetti-etal-1999}, and Ury{\=u} and
Eriguchi~\cite{uryu-eriguchi-2000}.  A full description of the
numerical methods used by Bonazzola can be found in
Ref.~\cite{gourgoulhon-etal-2000}.

\newpage

\section{Acknowledgments}
This work was supported by NSF grant PHY-9988581 to Wake Forest
University.  I am especially indebted to James York, Saul Teukolsky,
and Stuart Shapiro for their support over the years.  I would also
like to thank Andrew Abrahams, Thomas Baumgarte, and Mark Scheel
for innumerable helpful discussions.

\newpage


\begin{thebibliography}{100}

\bibitem{abrahams-etal-1992}\comment{article}
\comment{author}Abrahams, A.M., Bernstein, D., Hobill, D., Seidel, E., and
  Smarr, L., \comment{title}``Numerically generated black-hole spacetimes:
  {I}nteraction with gravitational waves'', \comment{journal}{\em Phys. Rev.
  D}, \comment{volume}{\bf 45}\comment{number}(10), \comment{pages}3544--3558,
  (\comment{month}May, \comment{year}1992). \keywords{Numerical relativity,
  Black holes, Gravitational radiation}

\bibitem{abrahams-evans-1993}\comment{article}
\comment{author}Abrahams, A.M., and Evans, C.R., \comment{title}``Critical
  Behavior and Scaling in Vacuum Axisymmetric Gravitational Collapse'',
  \comment{journal}{\em Phys. Rev. Lett.}, \comment{volume}{\bf
  70}\comment{number}(20), \comment{pages}2980--2983, (\comment{month}May,
  \comment{year}1993). \keywords{Gravitational Collapse, Critical phenomena,
  Gravitational radiation}

\bibitem{abrahams-price-1996}\comment{article}
\comment{author}Abrahams, A.M., and Price, R.H., \comment{title}``Black-hole
  collisions from {B}rill-{L}indquist initial data: {P}redictions of
  perturbation theory'', \comment{journal}{\em Phys. Rev. D},
  \comment{volume}{\bf 53}\comment{number}(4), \comment{pages}1972--1976,
  (\comment{month}February, \comment{year}1996). For a related online version
  see: \comment{author}A.M. Abrahams, et al., \comment{onlinetitle}``Black-hole
  collisions from {B}rill-{L}indquist initial data: predictions of perturbation
  theory'', (\comment{onlinemonth}September, \comment{onlineyear}1995),
  \comment{fileformat}[Online Los Alamos Archive Preprint]: cited on
  \comment{cited}18 July 2000,
  \comment{onlineaddress}http://xxx.lanl.gov/abs/gr-qc/9509020.
  \keywords{Numerical relativity, Perturbation theory, Black holes}

\bibitem{anderson-york-1998}\comment{article}
\comment{author}Anderson, A., and York, Jr., J.W., \comment{title}``Hamiltonian
  Time Evolution for General Relativity'', \comment{journal}{\em Phys. Rev.
  Lett.}, \comment{volume}{\bf 81}\comment{number}(6),
  \comment{pages}1154--1157, (\comment{month}August, \comment{year}1998). For a
  related online version see: \comment{author}A.~Anderson, et al.,
  \comment{onlinetitle}``Hamiltonian Time Evolution for General Relativity'',
  (\comment{onlinemonth}July, \comment{onlineyear}1998),
  \comment{fileformat}[Online Los Alamos Archive Preprint]: cited on
  \comment{cited}18 July 2000,
  \comment{onlineaddress}http://xxx.lanl.gov/abs/gr-qc/9807041.
  \keywords{Canonical methods, Hamiltonian systems, Cauchy problem}

\bibitem{arbona-etal-1998}\comment{article}
\comment{author}Arbona, A., Bona, C., Carot, J., Mas, L., Mass{\`o}, J., and
  Stella, J., \comment{title}``Stuffed black holes'', \comment{journal}{\em
  Phys. Rev. D}, \comment{volume}{\bf 57}\comment{number}(4),
  \comment{pages}2397--2402, (\comment{month}February, \comment{year}1998). For
  a related online version see: \comment{author}A.~Arbona, et al.,
  \comment{onlinetitle}``Stuffed Black Holes'', (\comment{onlinemonth}October,
  \comment{onlineyear}1997), \comment{fileformat}[Online Los Alamos Archive
  Preprint]: cited on \comment{cited}18 July 2000,
  \comment{onlineaddress}http://xxx.lanl.gov/abs/gr-qc/9710111.
  \keywords{Numerical relativity, Initial value problem, Black holes}

\bibitem{ADM}\comment{inbook}
\comment{author}Arnowitt, R., Deser, S., and Misner, C.W., \comment{title}``The
  dynamics of general relativity'', in \comment{editor}Witten, L., ed.,
  \comment{booktitle}{\em Gravitation: An Introduction to Current Research},
  \comment{pages} 227--265, (\comment{publisher}Wiley, \comment{address}New
  York, \comment{year}1962). \keywords{ADM formalism, Cauchy problem}

\bibitem{asada-1998}\comment{article}
\comment{author}Asada, H., \comment{title}``Formulation for the internal motion
  of quasiequilibrium configurations in general relativity'',
  \comment{journal}{\em Phys. Rev. D}, \comment{volume}{\bf
  57}\comment{number}(12), \comment{pages}7292--7298, (\comment{month}June,
  \comment{year}1998). For a related online version see:
  \comment{author}H.~Asada, \comment{onlinetitle}``Formulation for the internal
  motion of quasi-equilibrium configurations in general relativity'',
  (\comment{onlinemonth}April, \comment{onlineyear}1998),
  \comment{fileformat}[Online Los Alamos Archive Preprint]: cited on
  \comment{cited}18 July 2000,
  \comment{onlineaddress}http://xxx.lanl.gov/abs/gr-qc/9804003.
  \keywords{Neutron stars, Binary systems, Hydrodynamics}

\bibitem{bardeen-1970}\comment{article}
\comment{author}Bardeen, J.M., \comment{title}``A Variational Principle for
  Rotating Stars in General Relativity'', \comment{journal}{\em Astrophys. J.},
  \comment{volume}{\bf 162}\comment{number}(1), \comment{pages}71--95,
  (\comment{month}October, \comment{year}1970). \keywords{Relativistic stars}

\bibitem{bardeen-wagoner-1971}\comment{article}
\comment{author}Bardeen, J.M., and Wagoner, R.V., \comment{title}``Relativistic
  Disks. {I}. {U}niform Rotation'', \comment{journal}{\em Astrophys. J.},
  \comment{volume}{\bf 167}\comment{number}(3), \comment{pages}359--423,
  (\comment{month}August, \comment{year}1971). \keywords{Relativistic stars,
  Numerical methods}

\bibitem{baumgarte-2000}\comment{article}
\comment{author}Baumgarte, T.W., \comment{title}``Innermost stable circular
  orbit of binary black holes'', \comment{journal}{\em Phys. Rev. D},
  \comment{volume}{\bf 62}\comment{number}(2), \comment{pages}024018/1--8,
  (\comment{month}July, \comment{year}2000). For a related online version see:
  \comment{author}T.W. Baumgarte, \comment{onlinetitle}``The Innermost Stable
  Circular Orbit of Binary Black Holes'', (\comment{onlinemonth}April,
  \comment{onlineyear}2000), \comment{fileformat}[Online Los Alamos Archive
  Preprint]: cited on \comment{cited}18 July 2000,
  \comment{onlineaddress}http://xxx.lanl.gov/abs/gr-qc/0004050.
  \keywords{Numerical relativity, Binary systems, Black holes}

\bibitem{baumgarte-etal-1997}\comment{article}
\comment{author}Baumgarte, T.W., Cook, G.B., Scheel, M.A., Shapiro, S.L., and
  Teukolsky, S.A., \comment{title}``Binary Neutron Stars in General Relativity:
  {Q}uasi-Equilibrium Models'', \comment{journal}{\em Phys. Rev. Lett.},
  \comment{volume}{\bf 79}\comment{number}(7), \comment{pages}1182--1185,
  (\comment{month}August, \comment{year}1997). For a related online version
  see: \comment{author}T.W. Baumgarte, et al., \comment{onlinetitle}``Binary
  Neutron Stars in General Relativity: {Q}uasi-Equilibrium Models'',
  (\comment{onlinemonth}April, \comment{onlineyear}1997),
  \comment{fileformat}[Online Los Alamos Archive Preprint]: cited on
  \comment{cited}18 July 2000,
  \comment{onlineaddress}http://xxx.lanl.gov/abs/gr-qc/9704024.
  \keywords{Neutron Stars, Binary systems, Approximation methods}

\bibitem{baumgarte-etal-1998-model}\comment{article}
\comment{author}Baumgarte, T.W., Cook, G.B., Scheel, M.A., Shapiro, S.L., and
  Teukolsky, S.A., \comment{title}``General Relativistic Models of Binary
  Neutron Stars in Quasiequilibrium'', \comment{journal}{\em Phys. Rev. D},
  \comment{volume}{\bf 57}\comment{number}(12), \comment{pages}7299--7311,
  (\comment{month}June, \comment{year}1998). For a related online version see:
  \comment{author}T.W. Baumgarte, et al., \comment{onlinetitle}``General
  Relativistic Models of Binary Neutron Stars in Quasiequilibrium'',
  (\comment{onlinemonth}September, \comment{onlineyear}1997),
  \comment{fileformat}[Online Los Alamos Archive Preprint]: cited on
  \comment{cited}18 July 2000,
  \comment{onlineaddress}http://xxx.lanl.gov/abs/gr-qc/9709026.
  \keywords{Neutron Stars, Binary systems, Approximation methods}

\bibitem{baumgarte-etal-1998-stab}\comment{article}
\comment{author}Baumgarte, T.W., Cook, G.B., Scheel, M.A., Shapiro, S.L., and
  Teukolsky, S.A., \comment{title}``The Stability of Relativistic Neutron Stars
  in Binary Orbit'', \comment{journal}{\em Phys. Rev. D}, \comment{volume}{\bf
  57}\comment{number}(10), \comment{pages}6181--6184, (\comment{month}May,
  \comment{year}1998). For a related online version see: \comment{author}T.W.
  Baumgarte, et al., \comment{onlinetitle}``The Stability of Relativistic
  Neutron Stars in Binary Orbit'', (\comment{onlinemonth}May,
  \comment{onlineyear}1997), \comment{fileformat}[Online Los Alamos Archive
  Preprint]: cited on \comment{cited}18 July 2000,
  \comment{onlineaddress}http://xxx.lanl.gov/abs/gr-qc/9705023.
  \keywords{Neutron Stars, Binary systems, Stability theory}

\bibitem{bildsten-cutler-1992}\comment{article}
\comment{author}Bildsen, L., and Cutler, C., \comment{title}``Tidal Interaction
  of Inspiralling Compact Binaries'', \comment{journal}{\em Astrophys. J.},
  \comment{volume}{\bf 400}\comment{number}(1), \comment{pages}175--180,
  (\comment{month}November, \comment{year}1992). \keywords{Binary systems,
  Gravitational wave sources, Neutron stars, Black holes}

\bibitem{bishop-etal-1998}\comment{article}
\comment{author}Bishop, N.T., Isaacson, R., Maharaj, M., and Winicour, J.,
  \comment{title}``Black hole data via a {K}err-{S}child approach'',
  \comment{journal}{\em Phys. Rev. D}, \comment{volume}{\bf
  57}\comment{number}(10), \comment{pages}6113--6118, (\comment{month}May,
  \comment{year}1998). For a related online version see: \comment{author}N.T.
  Bishop, et al., \comment{onlinetitle}``Black Hole Data via a Kerr-Schild
  Approach'', (\comment{onlinemonth}November, \comment{onlineyear}1997),
  \comment{fileformat}[Online Los Alamos Archive Preprint]: cited on
  \comment{cited}18 July 2000,
  \comment{onlineaddress}http://xxx.lanl.gov/abs/gr-qc/9711076.
  \keywords{Constraint equations, Black holes, Binary systems}

\bibitem{bocquet-etal-1995}\comment{article}
\comment{author}Bocquet, M., Bonazzola, S., Gourgoulhon, E., and Novak, J.,
  \comment{title}``Rotating Neutron Star Models With Magnetic Field'',
  \comment{journal}{\em Astron. Astrophys.}, \comment{volume}{\bf
  301}\comment{number}(3), \comment{pages}757--775, (\comment{month}September,
  \comment{year}1995). For a related online version see:
  \comment{author}M.~Bocquet, et al., \comment{onlinetitle}``Rotating Neutron
  Star Models With Magnetic Field'', (\comment{onlinemonth}March,
  \comment{onlineyear}1995), \comment{fileformat}[Online Los Alamos Archive
  Preprint]: cited on \comment{cited}18 July 2000,
  \comment{onlineaddress}http://xxx.lanl.gov/abs/gr-qc/9503044.
  \keywords{Neutron stars, Numerical methods, Magnetic fields}

\bibitem{bona-masso-1988}\comment{article}
\comment{author}Bona, C., and Mass{\'o}, J., \comment{title}``Harmonic
  synchronizations of spacetime'', \comment{journal}{\em Phys. Rev. D},
  \comment{volume}{\bf 38}\comment{number}(8), \comment{pages}2419--2422,
  (\comment{month}October, \comment{year}1988). \keywords{Exact solutions,
  Black holes, Kerr-Newman metric}

\bibitem{bonazzola-etal-1997}\comment{article}
\comment{author}Bonazzola, S., Gourgoulhon, E., and Marck, J.-A.,
  \comment{title}``A relativistic formalism to compute quasi-equilibrium
  configurations of non-synchronized neutron star binaries'',
  \comment{journal}{\em Phys. Rev. D}, \comment{volume}{\bf
  56}\comment{number}(12), \comment{pages}7740--7749, (\comment{month}December,
  \comment{year}1997). For a related online version see:
  \comment{author}S.~Bonazzola, et al., \comment{onlinetitle}``A relativistic
  formalism to compute quasi-equilibrium configurations of non-synchronized
  neutron star binaries'', (\comment{onlinemonth}October,
  \comment{onlineyear}1997), \comment{fileformat}[Online Los Alamos Archive
  Preprint]: cited on \comment{cited}18 July 2000,
  \comment{onlineaddress}http://xxx.lanl.gov/abs/gr-qc/9710031.
  \keywords{Neutron stars, Binary systems, Hydrodynamics, Approximation
  methods}

\bibitem{bonazzola-etal-1998}\comment{article}
\comment{author}Bonazzola, S., Gourgoulhon, E., and Marck, J.-A.,
  \comment{title}``Numerical approach for high precision 3-{D} relativistic
  star models'', \comment{journal}{\em Phys. Rev. D}, \comment{volume}{\bf
  58}\comment{number}(10), \comment{pages}104020/1--14,
  (\comment{month}November, \comment{year}1998). For a related online version
  see: \comment{author}S.~Bonazzola, et al., \comment{onlinetitle}``Numerical
  approach for high precision 3-{D} relativistic star models'',
  (\comment{onlinemonth}March, \comment{onlineyear}1998),
  \comment{fileformat}[Online Los Alamos Archive Preprint]: cited on
  \comment{cited}18 July 2000,
  \comment{onlineaddress}http://xxx.lanl.gov/abs/astro-ph/9803086.
  \keywords{Numerical methods, Spectral methods}

\bibitem{bonazzola-etal-1999}\comment{article}
\comment{author}Bonazzola, S., Gourgoulhon, E., and Marck, J.-A.,
  \comment{title}``Numerical models of irrotational binary neutron stars in
  general relativity'', \comment{journal}{\em Phys. Rev. Lett.},
  \comment{volume}{\bf 82}\comment{number}(5), \comment{pages}892--895,
  (\comment{month}February, \comment{year}1999). For a related online version
  see: \comment{author}S.~Bonazzola, et al., \comment{onlinetitle}``Numerical
  models of irrotational binary neutron stars in general relativity'',
  (\comment{onlinemonth}October, \comment{onlineyear}1998),
  \comment{fileformat}[Online Los Alamos Archive Preprint]: cited on
  \comment{cited}18 July 2000,
  \comment{onlineaddress}http://xxx.lanl.gov/abs/gr-qc/9810072.
  \keywords{Neutron stars, Binary systems, Numerical methods}

\bibitem{bonazzola-etal-1993}\comment{article}
\comment{author}Bonazzola, S., Gourgoulhon, E., Salgado, M., and Marck, J.-A.,
  \comment{title}``Axisymmetric rotating relativistic bodies: a new numerical
  approach for `exact' solutions'', \comment{journal}{\em Astron. Astrophys.},
  \comment{volume}{\bf 278}\comment{number}(2), \comment{pages}421--443,
  (\comment{month}November, \comment{year}1993). \keywords{Neutron stars,
  Numerical methods, Spectral methods}

\bibitem{bonazzola-schneider-1974}\comment{article}
\comment{author}Bonazzola, S., and Schneider, J., \comment{title}``An Exact
  Study of Rigidly and Rapidly Rotating Stars in General Relativity with
  Application to the Crab Pulsar'', \comment{journal}{\em Astrophys. J.},
  \comment{volume}{\bf 191}\comment{number}(1), \comment{pages}273--286,
  (\comment{month}July, \comment{year}1974). \keywords{Neutron stars, Numerical
  methods}

\bibitem{bowen-1979}\comment{article}
\comment{author}Bowen, J.M., \comment{title}``General form for the longitudinal
  momentum of a spherically symmetric source'', \comment{journal}{\em Gen.
  Relativ. Gravit.}, \comment{volume}{\bf 11}\comment{number}(3),
  \comment{pages}227--231, (\comment{month}October, \comment{year}1979).
  \keywords{Initial value problem, Constraint equations}

\bibitem{bowen-1982}\comment{article}
\comment{author}Bowen, J.M., \comment{title}``General solution for flat-space
  longitudinal momentum'', \comment{journal}{\em Gen. Relativ. Gravit.},
  \comment{volume}{\bf 14}\comment{number}(12), \comment{pages}1183--1191,
  (\comment{month}December, \comment{year}1982). \keywords{Initial value
  problem, Constraint equations}

\bibitem{bowen-etal-1984}\comment{article}
\comment{author}Bowen, J.M., Rauber, J., and York, Jr., J.W.,
  \comment{title}``Two black holes with axisymmetric parallel spins: {I}nitial
  data'', \comment{journal}{\em Class. Quantum Grav.}, \comment{volume}{\bf
  1}\comment{number}(5), \comment{pages}591--610, (\comment{month}September,
  \comment{year}1984). \keywords{Initial value problem, Constraint equations,
  Black holes}

\bibitem{bowen-york-1980}\comment{article}
\comment{author}Bowen, J.M., and York, Jr., J.W.,
  \comment{title}``Time-asymmetric initial data for black holes and black-hole
  collisions'', \comment{journal}{\em Phys. Rev. D}, \comment{volume}{\bf
  21}\comment{number}(8), \comment{pages}2047--2056, (\comment{month}April,
  \comment{year}1980). \keywords{Initial value problem, Constraint equations,
  Black holes}

\bibitem{brandt-bruegmann-1997}\comment{article}
\comment{author}Brandt, S., and Br{\"u}gmann, B., \comment{title}``A simple
  construction of initial data for multiple black holes'',
  \comment{journal}{\em Phys. Rev. Lett.}, \comment{volume}{\bf
  78}\comment{number}(19), \comment{pages}3606--3609, (\comment{month}May,
  \comment{year}1997). For a related online version see:
  \comment{author}S.~Brandt, et al., \comment{onlinetitle}``A simple
  construction of initial data for multiple black holes'',
  (\comment{onlinemonth}March, \comment{onlineyear}1997),
  \comment{fileformat}[Online Los Alamos Archive Preprint]: cited on
  \comment{cited}18 July 2000,
  \comment{onlineaddress}http://xxx.lanl.gov/abs/gr-qc/9703066.
  \keywords{Initial value problem, Constraint equations, Black holes}

\bibitem{brandt-seidel-1995-I}\comment{article}
\comment{author}Brandt, S.R., and Seidel, E., \comment{title}``Evolution of
  distorted rotating black holes. {I}: {M}ethods and tests'',
  \comment{journal}{\em Phys. Rev. D}, \comment{volume}{\bf
  52}\comment{number}(2), \comment{pages}856--869, (\comment{month}July,
  \comment{year}1995). For a related online version see: \comment{author}S.R.
  Brandt, et al., \comment{onlinetitle}``The Evolution of Distorted Rotating
  Black Holes I: Methods and Tests'', (\comment{onlinemonth}December,
  \comment{onlineyear}1994), \comment{fileformat}[Online Los Alamos Archive
  Preprint]: cited on \comment{cited}18 July 2000,
  \comment{onlineaddress}http://xxx.lanl.gov/abs/gr-qc/9412072.
  \keywords{Numerical relativity, Black holes, Gravitational radiation}

\bibitem{brandt-seidel-1995-II}\comment{article}
\comment{author}Brandt, S.R., and Seidel, E., \comment{title}``Evolution of
  distorted rotating black holes. {II}: {D}ynamics and analysis'',
  \comment{journal}{\em Phys. Rev. D}, \comment{volume}{\bf
  52}\comment{number}(2), \comment{pages}870--886, (\comment{month}July,
  \comment{year}1995). For a related online version see: \comment{author}S.R.
  Brandt, et al., \comment{onlinetitle}``The Evolution of Distorted Rotating
  Black Holes II: Dynamics and Analysis'', (\comment{onlinemonth}December,
  \comment{onlineyear}1994), \comment{fileformat}[Online Los Alamos Archive
  Preprint]: cited on \comment{cited}18 July 2000,
  \comment{onlineaddress}http://xxx.lanl.gov/abs/gr-qc/9412073.
  \keywords{Numerical relativity, Black holes, Gravitational radiation}

\bibitem{brandt-seidel-1996}\comment{article}
\comment{author}Brandt, S.R., and Seidel, E., \comment{title}``Evolution of
  distorted rotating black holes. {III}: {I}nitial data'',
  \comment{journal}{\em Phys. Rev. D}, \comment{volume}{\bf
  54}\comment{number}(2), \comment{pages}1403--1416, (\comment{month}July,
  \comment{year}1996). For a related online version see: \comment{author}S.R.
  Brandt, et al., \comment{onlinetitle}``The Evolution of Distorted Rotating
  Black Holes III: Initial Data'', (\comment{onlinemonth}January,
  \comment{onlineyear}1996), \comment{fileformat}[Online Los Alamos Archive
  Preprint]: cited on \comment{cited}18 July 2000,
  \comment{onlineaddress}http://xxx.lanl.gov/abs/gr-qc/9601010.
  \keywords{Numerical relativity, Black holes, Initial value problem}

\bibitem{brill-lindquist-1963}\comment{article}
\comment{author}Brill, D.R., and Lindquist, R.W., \comment{title}``Interaction
  energy in geometrostatics'', \comment{journal}{\em Phys. Rev.},
  \comment{volume}{\bf 131}\comment{number}(1), \comment{pages}471--476,
  (\comment{month}July, \comment{year}1963). \keywords{Initial value problem,
  Black holes}

\bibitem{butterworth-1976}\comment{article}
\comment{author}Butterworth, E.M., \comment{title}``On the Structure and
  Stability of Rapidly Rotating Fluid Bodies in General Relativity. {II}. {T}he
  Structure of Uniformly Rotating Pseudopolytropes'', \comment{journal}{\em
  Astrophys. J.}, \comment{volume}{\bf 204}\comment{number}(2),
  \comment{pages}561--572, (\comment{month}March, \comment{year}1976).
  \keywords{Neutron stars, Numerical methods}

\bibitem{butterworth-1979}\comment{article}
\comment{author}Butterworth, E.M., \comment{title}``On the Structure and
  Stability of Rapidly Rotating Fluid Bodies in General Relativity. {III}.
  {B}eyond the Angular Velocity Peak'', \comment{journal}{\em Astrophys. J.},
  \comment{volume}{\bf 231}\comment{number}(1), \comment{pages}219--223,
  (\comment{month}July, \comment{year}1979). \keywords{Neutron stars, Numerical
  methods}

\bibitem{butterworth-ipser-1975}\comment{article}
\comment{author}Butterworth, E.M., and Ipser, J.R., \comment{title}``Rapidly
  Rotating Fluid Bodies in General Relativity'', \comment{journal}{\em
  Astrophys. J. Lett.}, \comment{volume}{\bf 200}\comment{number}(2),
  \comment{pages}L103--L106, (\comment{month}September, \comment{year}1975).
  \keywords{Neutron stars, Numerical methods}

\bibitem{butterworth-ipser-1976}\comment{article}
\comment{author}Butterworth, E.M., and Ipser, J.R., \comment{title}``On the
  Structure and Stability of Rapidly Rotating Fluid Bodies in General
  Relativity. {I}. {T}he Numerical Method for Computing Structure and Its
  Application to Uniformly Rotating Homogeneous Bodies'', \comment{journal}{\em
  Astrophys. J.}, \comment{volume}{\bf 204}\comment{number}(1),
  \comment{pages}200--233, (\comment{month}February, \comment{year}1976).
  \keywords{Neutron stars, Numerical methods}

\bibitem{cantor-kulkarni-1982}\comment{article}
\comment{author}Cantor, M., and Kulkarni, A.D., \comment{title}``Physical
  distinctions between normalized solutions of the two-body problem of general
  relativity'', \comment{journal}{\em Phys. Rev. D}, \comment{volume}{\bf
  25}\comment{number}(10), \comment{pages}2521--2526, (\comment{month}May,
  \comment{year}1982). \keywords{Initial value problem, Black holes}

\bibitem{choquet-york-GRG}\comment{inbook}
\comment{author}Choquet-Bruhat, Y., and York, Jr., J.W., \comment{title}``The
  {C}auchy problem'', in \comment{editor}Held, A., ed., \comment{booktitle}{\em
  General Relativity and Gravitation: {O}ne Hundred Years After the Birth of
  {A}lbert {E}instein}, volume~1, \comment{pages} 99--172,
  (\comment{publisher}Plenum, \comment{address}New York, \comment{year}1980).
  \keywords{ADM formalism, Cauchy problem}

\bibitem{cook-1991}\comment{article}
\comment{author}Cook, G.B., \comment{title}``Initial data for axisymmetric
  black-hole collisions'', \comment{journal}{\em Phys. Rev. D},
  \comment{volume}{\bf 44}\comment{number}(10), \comment{pages}2983--3000,
  (\comment{month}November, \comment{year}1991). \keywords{Numerical
  relativity, Constraint equations, Black holes}

\bibitem{cook-1994}\comment{article}
\comment{author}Cook, G.B., \comment{title}``Three-dimensional initial data for
  the collision of two black holes {II}: {Q}uasicircular orbits for equal mass
  black holes'', \comment{journal}{\em Phys. Rev. D}, \comment{volume}{\bf
  50}\comment{number}(8), \comment{pages}5025--5032, (\comment{month}October,
  \comment{year}1994). For a related online version see: \comment{author}G.B.
  Cook, \comment{onlinetitle}``Three-dimensional initial data for the collision
  of two black holes {II:} {Quasicircular} orbits for equal mass black holes'',
  (\comment{onlinemonth}April, \comment{onlineyear}1994),
  \comment{fileformat}[Online Los Alamos Archive Preprint]: cited on
  \comment{cited}18 July 2000,
  \comment{onlineaddress}http://xxx.lanl.gov/abs/gr-qc/9404043.
  \keywords{Numerical relativity, Binary systems, Black holes}

\bibitem{cook-abrahams-1992}\comment{article}
\comment{author}Cook, G.B., and Abrahams, A.M., \comment{title}``Horizon
  structure of initial-data sets for axisymmetric two-black-hole collisions'',
  \comment{journal}{\em Phys. Rev. D}, \comment{volume}{\bf
  46}\comment{number}(2), \comment{pages}702--713, (\comment{month}July,
  \comment{year}1992). \keywords{Numerical relativity, Constraint equations,
  Black holes, Perturbation methods}

\bibitem{cook-etal-1993}\comment{article}
\comment{author}Cook, G.B., Choptuik, M.W., Dubal, M.R., Klasky, S., Matzner,
  R.A., and Oliveira, S.R., \comment{title}``Three-dimensional initial data for
  the collision of two black holes'', \comment{journal}{\em Phys. Rev. D},
  \comment{volume}{\bf 47}\comment{number}(4), \comment{pages}1471--1490,
  (\comment{month}February, \comment{year}1993). \keywords{Numerical
  relativity, Constraint equations, Black holes, Numerical methods}

\bibitem{cook-scheel-1997}\comment{article}
\comment{author}Cook, G.B., and Scheel, M.A., \comment{title}``Well-behaved
  harmonic time slices of a charged, rotating, boosted black hole'',
  \comment{journal}{\em Phys. Rev. D}, \comment{volume}{\bf
  56}\comment{number}(8), \comment{pages}4775--4781, (\comment{month}October,
  \comment{year}1997). \keywords{Exact solutions, Black holes, Kerr-Newman
  metric}

\bibitem{cook-etal-1992}\comment{article}
\comment{author}Cook, G.B., Shapiro, S.L., and Teukolsky, S.A.,
  \comment{title}``Spin-up of a rapidly rotating star by angular momentum loss:
  {E}ffects of general relativity'', \comment{journal}{\em Astrophys. J.},
  \comment{volume}{\bf 398}\comment{number}(1), \comment{pages}203--223,
  (\comment{month}October, \comment{year}1992). \keywords{Neutron stars,
  Numerical methods}

\bibitem{cook-etal-1994-R}\comment{article}
\comment{author}Cook, G.B., Shapiro, S.L., and Teukolsky, S.A.,
  \comment{title}``Rapidly rotating neutron stars in general relativity:
  {R}ealistic equations of state'', \comment{journal}{\em Astrophys. J.},
  \comment{volume}{\bf 424}\comment{number}(2), \comment{pages}823--845,
  (\comment{month}April, \comment{year}1994). \keywords{Neutron stars,
  Numerical methods}

\bibitem{cook-etal-1994-P}\comment{article}
\comment{author}Cook, G.B., Shapiro, S.L., and Teukolsky, S.A.,
  \comment{title}``Rapidly rotating polytropes in general relativity'',
  \comment{journal}{\em Astrophys. J.}, \comment{volume}{\bf
  422}\comment{number}(1), \comment{pages}227--242, (\comment{month}February,
  \comment{year}1994). \keywords{Neutron stars, Numerical methods}

\bibitem{cook-etal-1996}\comment{article}
\comment{author}Cook, G.B., Shapiro, S.L., and Teukolsky, S.A.,
  \comment{title}``Testing a Simplified Version of {E}instein's Equations for
  Numerical Relativity'', \comment{journal}{\em Phys. Rev. D},
  \comment{volume}{\bf 53}\comment{number}(10), \comment{pages}5533--5540,
  (\comment{month}May, \comment{year}1996). For a related online version see:
  \comment{author}G.B. Cook, et al., \comment{onlinetitle}``Testing a
  Simplified Version of {E}instein's Equations for Numerical Relativity'',
  (\comment{onlinemonth}December, \comment{onlineyear}1995),
  \comment{fileformat}[Online Los Alamos Archive Preprint]: cited on
  \comment{cited}18 July 2000,
  \comment{onlineaddress}http://xxx.lanl.gov/abs/gr-qc/9512009.
  \keywords{Constraint equations, Approximation methods, Neutron stars}

\bibitem{Damour-etal-2000}\comment{online}
\comment{author}Damour, T., Jaranowski, P., and Schaefer, G.,
  \comment{onlinetitle}``On the determination of the last stable orbit for
  circular general relativistic binaries at the third post-{N}ewtonian
  approximation'', (\comment{onlinemonth}May, \comment{onlineyear}2000),
  \comment{fileformat}[Online Los Alamos Archive Preprint]: cited on
  \comment{cited}18 July 2000,
  \comment{onlineaddress}http://xxx.lanl.gov/abs/gr-qc/0005034. Submitted to
  {\it Phys. Rev. D}. \keywords{Post-Newtonian approximations, Binary systems}

\bibitem{doran-2000}\comment{article}
\comment{author}Doran, C., \comment{title}``A new form of the {K}err
  solution'', \comment{journal}{\em Phys. Rev. D}, \comment{volume}{\bf
  61}\comment{number}(6), \comment{pages}067503/1--4, (\comment{month}March,
  \comment{year}2000). For a related online version see:
  \comment{author}C.~Doran, \comment{onlinetitle}``A new form of the {Kerr}
  solution'', (\comment{onlinemonth}October, \comment{onlineyear}1999),
  \comment{fileformat}[Online Los Alamos Archive Preprint]: cited on
  \comment{cited}18 July 2000,
  \comment{onlineaddress}http://xxx.lanl.gov/abs/gr-qc/9910099. \keywords{Exact
  solutions, Black holes, Kerr metric}

\bibitem{Frontiers}\comment{proceedings}
\comment{editor}Evans, C.R., Finn, L.S., and Hobill, D.W., eds.,
  \comment{title}{\em Frontiers in Numerical Relativity},
  (\comment{publisher}Cambridge University Press, \comment{address}Cambridge,
  England, \comment{year}1989). \keywords{}

\bibitem{flanagan-1999}\comment{article}
\comment{author}Flanagan, E.E., \comment{title}``Possible Explanation for
  Star-Crushing Effect in Binary Neutron Star Simulations'',
  \comment{journal}{\em Phys. Rev. Lett.}, \comment{volume}{\bf
  82}\comment{number}(7), \comment{pages}1354--1357, (\comment{month}February,
  \comment{year}1999). For a related online version see: \comment{author}E.E.
  Flanagan, \comment{onlinetitle}``Possible explanation for star-crushing
  effect in binary neutron star simulations'', (\comment{onlinemonth}November,
  \comment{onlineyear}1998), \comment{fileformat}[Online Los Alamos Archive
  Preprint]: cited on \comment{cited}18 July 2000,
  \comment{onlineaddress}http://xxx.lanl.gov/abs/astro-ph/9811132.
  \keywords{Neutron stars, Binary systems, Gravitational collapse}

\bibitem{friedman-ipser-1992-E}\comment{article}
\comment{author}Friedman, J.L., and Ipser, J.R., \comment{title}``Errata:
  ``{R}apidly rotating relativistic stars'''', \comment{journal}{\em Philos.
  Trans. Roy. Soc. London, Ser. A}, \comment{volume}{\bf
  341}\comment{number}(1662), \comment{pages}561, (\comment{month}December,
  \comment{year}1992). \keywords{Neutron stars, Numerical methods}

\bibitem{friedman-ipser-1992}\comment{article}
\comment{author}Friedman, J.L., and Ipser, J.R., \comment{title}``Rapidly
  rotating relativistic stars'', \comment{journal}{\em Philos. Trans. Roy. Soc.
  London, Ser. A}, \comment{volume}{\bf 340}\comment{number}(1658),
  \comment{pages}391--422, (\comment{month}September, \comment{year}1992).
  \keywords{Neutron stars, Numerical methods}

\bibitem{Friedman-etal-1986}\comment{article}
\comment{author}Friedman, J.L., Ipser, J.R., and Parker, L.,
  \comment{title}``Rapidly Rotating Neutron Star Models'',
  \comment{journal}{\em Astrophys. J.}, \comment{volume}{\bf
  304}\comment{number}(1), \comment{pages}115--139, (\comment{month}May,
  \comment{year}1986). \keywords{Neutron stars, Numerical methods}

\bibitem{friedman-etal-1988}\comment{article}
\comment{author}Friedman, J.L., Ipser, J.R., and Sorkin, R.D.,
  \comment{title}``Turning-Point Method for Axisymmetric Stability of Rotating
  Relativistic Stars'', \comment{journal}{\em Astrophys. J.},
  \comment{volume}{\bf 325}\comment{number}(2), \comment{pages}722--274,
  (\comment{month}February, \comment{year}1988). \keywords{Neutron stars,
  Stability theory}

\bibitem{garat-price-2000}\comment{article}
\comment{author}Garat, A., and Price, R.H., \comment{title}``Nonexistence of
  conformally flat slices of the {K}err spacetime'', \comment{journal}{\em
  Phys. Rev. D}, \comment{volume}{\bf 61}\comment{number}(12),
  \comment{pages}124011/1--4, (\comment{month}June, \comment{year}2000). For a
  related online version see: \comment{author}A.~Garat, et al.,
  \comment{onlinetitle}``Nonexistence of conformally flat slices of the Kerr
  spacetime'', (\comment{onlinemonth}February, \comment{onlineyear}2000),
  \comment{fileformat}[Online Los Alamos Archive Preprint]: cited on
  \comment{cited}18 July 2000,
  \comment{onlineaddress}http://xxx.lanl.gov/abs/gr-qc/0002013. \keywords{Kerr
  metric, Existence theorems}

\bibitem{gourgoulhon-1998}\comment{online}
\comment{author}Gourgoulhon, E., \comment{onlinetitle}``Relations between three
  formalisms for irrotational binary neutron stars in general relativity'',
  (\comment{onlinemonth}April, \comment{onlineyear}1998),
  \comment{fileformat}[Online Los Alamos Archive Preprint]: cited on
  \comment{cited}18 July 2000,
  \comment{onlineaddress}http://xxx.lanl.gov/abs/gr-qc/9804054.
  \keywords{Neutron stars, Binary systems, Hydrodynamics}

\bibitem{gourgoulhon-bonazzola-1993}\comment{article}
\comment{author}Gourgoulhon, E., and Bonazzola, S.,
  \comment{title}``Noncircular axisymmetric stationary spacetimes'',
  \comment{journal}{\em Phys. Rev. D}, \comment{volume}{\bf
  48}\comment{number}(6), \comment{pages}2635--2652, (\comment{month}September,
  \comment{year}1993). \keywords{Neutron stars, Numerical methods}

\bibitem{gourgoulhon-etal-2000}\comment{online}
\comment{author}Gourgoulhon, E., Grandcl{\`e}ment, P., Taniguchi, K., Marck,
  J.-A., and Bonazzola, S., \comment{onlinetitle}``Quasiequilibrium sequences
  of synchronized and irrotational binary neutron stars in general relativity.
  I. Method and tests'', (\comment{onlinemonth}July, \comment{onlineyear}2000),
  \comment{fileformat}[Online Los Alamos Archive Preprint]: cited on
  \comment{cited}18 July 2000,
  \comment{onlineaddress}http://xxx.lanl.gov/abs/gr-qc/0007028. Submitted to
  {\it Phys. Rev. D}. \keywords{Neutron stars, Binary systems, Hydrodynamics,
  Approximation methods}

\bibitem{gullstrand-1922}\comment{article}
\comment{author}Gullstrand, A., \comment{journal}{\em Arkiv. Mat. Astron.
  Fys.}, \comment{volume}{\bf 16}, \comment{pages}1, (\comment{year}1922).
  \keywords{Exact solutions, Black holes, Schwarzschild metric}

\bibitem{kley-schaefer-1999}\comment{article}
\comment{author}Kley, W., and Sch{\"a}fer, G., \comment{title}``Relativistic
  dust disks and the {W}ilson-{M}athews approach'', \comment{journal}{\em Phys.
  Rev. D}, \comment{volume}{\bf 60}\comment{number}(2),
  \comment{pages}027501/1--4, (\comment{month}July, \comment{year}1999). For a
  related online version see: \comment{author}W.~Kley, et al.,
  \comment{onlinetitle}``Relativistic dust disks and the {W}ilson-{M}athews
  approach'', (\comment{onlinemonth}December, \comment{onlineyear}1998),
  \comment{fileformat}[Online Los Alamos Archive Preprint]: cited on
  \comment{cited}18 July 2000,
  \comment{onlineaddress}http://xxx.lanl.gov/abs/gr-qc/9812068.
  \keywords{Numerical Relativity, Approximation methods}

\bibitem{kochanek-1992}\comment{article}
\comment{author}Kochanek, C.S., \comment{title}``Coalescing Binary Neutron
  Stars'', \comment{journal}{\em Astrophys. J.}, \comment{volume}{\bf
  398}\comment{number}(1), \comment{pages}234--247, (\comment{month}October,
  \comment{year}1992). \keywords{Binary systems, Gravitational wave sources,
  Neutron stars}

\bibitem{komatsu-etal-1989-I}\comment{article}
\comment{author}Komatsu, H., Eriguchi, Y., and Hachisu, I.,
  \comment{title}``Rapidly rotating general relativistic stars--{I}.
  {N}umerical method and its application to uniformly rotating polytropes'',
  \comment{journal}{\em Mon. Not. R. Astr. Soc.}, \comment{volume}{\bf
  237}\comment{number}(2), \comment{pages}355--379, (\comment{month}March,
  \comment{year}1989). \keywords{Neutron stars, Numerical methods}

\bibitem{komatsu-etal-1989-II}\comment{article}
\comment{author}Komatsu, H., Eriguchi, Y., and Hachisu, I.,
  \comment{title}``Rapidly rotating general relativistic stars--{II}.
  {D}ifferentially rotating polytropes'', \comment{journal}{\em Mon. Not. R.
  Astr. Soc.}, \comment{volume}{\bf 239}\comment{number}(1),
  \comment{pages}153--171, (\comment{month}July, \comment{year}1989).
  \keywords{Neutron stars, Numerical methods}

\bibitem{kraus-wilczek-1994}\comment{article}
\comment{author}Kraus, P., and Wilczek, F., \comment{title}``Some applications
  of a simple stationary line element for the {S}chwarzschild geometry'',
  \comment{journal}{\em Mod. Phys. Lett. A}, \comment{volume}{\bf
  9}\comment{number}(40), \comment{pages}3713--3719, (\comment{month}December,
  \comment{year}1994). For a related online version see:
  \comment{author}P.~Kraus, et al., \comment{onlinetitle}``A simple stationary
  line element for the {Schwarzschild} geometry, and some applications'',
  (\comment{onlinemonth}June, \comment{onlineyear}1994),
  \comment{fileformat}[Online Los Alamos Archive Preprint]: cited on
  \comment{cited}18 July 2000,
  \comment{onlineaddress}http://xxx.lanl.gov/abs/gr-qc/9406042. \keywords{Exact
  solutions, Black holes, Reissner-Nordstrom solution}

\bibitem{kulkarni-1984}\comment{article}
\comment{author}Kulkarni, A.D., \comment{title}``Time-asymmetric initial data
  for the {$N$} black hole problem in general relativity'',
  \comment{journal}{\em J. Math. Phys.}, \comment{volume}{\bf
  25}\comment{number}(4), \comment{pages}1028--1034, (\comment{month}April,
  \comment{year}1984). \keywords{Initial value problem, Constraint equations,
  Black holes}

\bibitem{kulkarni-etal-1983}\comment{article}
\comment{author}Kulkarni, A.D., Shepley, L.C., and York, Jr., J.W.,
  \comment{title}``Initial data for {$N$} black holes'', \comment{journal}{\em
  Phys. Lett. A}, \comment{volume}{\bf 96}\comment{number}(5),
  \comment{pages}228--230, (\comment{month}July, \comment{year}1983).
  \keywords{Initial value problem, Constraint equations, Black holes}

\bibitem{lake-1994}\comment{online}
\comment{author}Lake, K., \comment{onlinetitle}``A class of quasi-stationary
  regular line elements for the {S}chwarzschild geometry'',
  (\comment{onlinemonth}July, \comment{onlineyear}1994),
  \comment{fileformat}[Online Los Alamos Archive Preprint]: cited on
  \comment{cited}18 July 2000,
  \comment{onlineaddress}http://xxx.lanl.gov/abs/gr-qc/9407005. \keywords{Exact
  solutions, Black holes, Schwarzschild metric}

\bibitem{lichnerowicz-1944}\comment{article}
\comment{author}Lichnerowicz, A., \comment{title}``L'integration des
  {\'e}quations de la gravitation relativiste et le probl{\`e}me des $n$
  corps'', \comment{journal}{\em J. Math. Pures et Appl.}, \comment{volume}{\bf
  23}, \comment{pages}37--63, (\comment{year}1944). \keywords{Cauchy problem,
  Initial value problem, Constraint equations}

\bibitem{lindquist-1963}\comment{article}
\comment{author}Lindquist, R.W., \comment{title}``Initial-Value Problem on
  {E}instein-{R}osen Manifolds'', \comment{journal}{\em J. Math. Phys.},
  \comment{volume}{\bf 4}\comment{number}(7), \comment{pages}938--950,
  (\comment{month}July, \comment{year}1963). \keywords{Initial value problem,
  Black holes}

\bibitem{lousto-price-1998}\comment{article}
\comment{author}Lousto, C.O., and Price, R.H., \comment{title}``Improved
  initial data for black hole collisions'', \comment{journal}{\em Phys. Rev.
  D}, \comment{volume}{\bf 57}\comment{number}(2), \comment{pages}1073--1083,
  (\comment{month}June, \comment{year}1998). For a related online version see:
  \comment{author}C.O. Lousto, et al., \comment{onlinetitle}``Improved initial
  data for black hole collisions'', (\comment{onlinemonth}August,
  \comment{onlineyear}1997), \comment{fileformat}[Online Los Alamos Archive
  Preprint]: cited on \comment{cited}18 July 2000,
  \comment{onlineaddress}http://xxx.lanl.gov/abs/gr-qc/9708022.
  \keywords{Numerical relativity, Black holes, Perturbation methods}

\bibitem{marronetti-etal-1998}\comment{article}
\comment{author}Marronetti, P., Mathews, G.J., and Wilson, J.R.,
  \comment{title}``Binary neutron-star systems: {F}rom the {N}ewtonian regime
  to the last stable orbit'', \comment{journal}{\em Phys. Rev. D},
  \comment{volume}{\bf 58}\comment{number}(10), \comment{pages}107503/1--4,
  (\comment{month}November, \comment{year}1998). For a related online version
  see: \comment{author}P.~Marronetti, et al., \comment{onlinetitle}``Binary
  Neutron-Star Systems: {F}rom the {N}ewtonian Regime to the Last Stable
  Orbit'', (\comment{onlinemonth}March, \comment{onlineyear}1998),
  \comment{fileformat}[Online Los Alamos Archive Preprint]: cited on
  \comment{cited}18 July 2000,
  \comment{onlineaddress}http://xxx.lanl.gov/abs/gr-qc/9803093.
  \keywords{Neutron stars, Binary systems, Approximation methods}

\bibitem{marronetti-etal-1999}\comment{article}
\comment{author}Marronetti, P., Mathews, G.J., and Wilson, J.R.,
  \comment{title}``Irrotational binary neutron stars in quasiequilibrium'',
  \comment{journal}{\em Phys. Rev. D}, \comment{volume}{\bf
  60}\comment{number}(8), \comment{pages}087301/1--4, (\comment{month}October,
  \comment{year}1999). For a related online version see:
  \comment{author}P.~Marronetti, et al., \comment{onlinetitle}``Irrotational
  binary neutron stars in quasiequilibrium'', (\comment{onlinemonth}June,
  \comment{onlineyear}1999), \comment{fileformat}[Online Los Alamos Archive
  Preprint]: cited on \comment{cited}18 July 2000,
  \comment{onlineaddress}http://xxx.lanl.gov/abs/gr-qc/9906088.
  \keywords{Neutron stars, Binary systems, Approximation methods}

\bibitem{marsa-choptuik-1996}\comment{article}
\comment{author}Marsa, R.L., and Choptuik, M.W.,
  \comment{title}``Black-hole-scalar-field interactions in spherical
  symmetry'', \comment{journal}{\em Phys. Rev. D}, \comment{volume}{\bf
  54}\comment{number}(8), \comment{pages}4929--4943, (\comment{month}October,
  \comment{year}1996). For a related online version see: \comment{author}R.L.
  Marsa, et al., \comment{onlinetitle}``Black Hole--Scalar Field Interactions
  in Spherical Symmetry'', (\comment{onlinemonth}July,
  \comment{onlineyear}1996), \comment{fileformat}[Online Los Alamos Archive
  Preprint]: cited on \comment{cited}18 July 2000,
  \comment{onlineaddress}http://xxx.lanl.gov/abs/gr-qc/9607034.
  \keywords{Numerical relativity, Black holes, Schwarzschild metric}

\bibitem{martel-poisson-2000}\comment{online}
\comment{author}Martel, K., and Poisson, E., \comment{onlinetitle}``Regular
  coordinate systems for {S}chwarzschild and other spherical spacetimes'',
  (\comment{onlinemonth}January, \comment{onlineyear}2000),
  \comment{fileformat}[Online Los Alamos Archive Preprint]: cited on
  \comment{cited}18 July 2000,
  \comment{onlineaddress}http://xxx.lanl.gov/abs/gr-qc/0001069. \keywords{Exact
  solutions, Black holes, Schwarzschild metric}

\bibitem{mathews-etal-1998}\comment{article}
\comment{author}Mathews, G.J., Marronetti, P., and Wilson, J.R.,
  \comment{title}``Relativistic hydrodynamics in close binary systems:
  {A}nalysis of neutron-star collapse'', \comment{journal}{\em Phys. Rev. D},
  \comment{volume}{\bf 58}\comment{number}(4), \comment{pages}043003/1--13,
  (\comment{month}August, \comment{year}1998). For a related online version
  see: \comment{author}G.J. Mathews, et al.,
  \comment{onlinetitle}``Relativistic Hydrodynamics in Close Binary Systems:
  {A}nalysis of Neutron-Star Collapse'', (\comment{onlinemonth}October,
  \comment{onlineyear}1997), \comment{fileformat}[Online Los Alamos Archive
  Preprint]: cited on \comment{cited}18 July 2000,
  \comment{onlineaddress}http://xxx.lanl.gov/abs/gr-qc/9710140.
  \keywords{Neutron stars, Binary systems, Gravitational collapse}

\bibitem{mathews-wilson-2000}\comment{article}
\comment{author}Mathews, G.J., and Wilson, J.R., \comment{title}``Revised
  relativistic hydrodynamical model for neutron-star binaries'',
  \comment{journal}{\em Phys. Rev. D}, \comment{volume}{\bf
  61}\comment{number}(12), \comment{pages}127304/1--4, (\comment{month}June,
  \comment{year}2000). For a related online version see: \comment{author}G.J.
  Mathews, et al., \comment{onlinetitle}``Revised Relativistic Hydrodynamical
  Model for Neutron-Star Binaries'', (\comment{onlinemonth}November,
  \comment{onlineyear}1999), \comment{fileformat}[Online Los Alamos Archive
  Preprint]: cited on \comment{cited}18 July 2000,
  \comment{onlineaddress}http://xxx.lanl.gov/abs/gr-qc/9911047.
  \keywords{Neutron stars, Binary systems, Approximation methods}

\bibitem{matzner-etal-1999}\comment{article}
\comment{author}Matzner, R.A., Huq, M.F., and Shoemaker, D.,
  \comment{title}``Initial Data and Coordinates for Multiple Black Hole
  Systems'', \comment{journal}{\em Phys. Rev. D}, \comment{volume}{\bf
  59}\comment{number}(2), \comment{pages}024015/1--6, (\comment{month}January,
  \comment{year}1999). For a related online version see: \comment{author}R.A.
  Matzner, et al., \comment{onlinetitle}``Initial Data and Coordinates for
  Multiple Black Hole Systems'', (\comment{onlinemonth}May,
  \comment{onlineyear}1998), \comment{fileformat}[Online Los Alamos Archive
  Preprint]: cited on \comment{cited}18 July 2000,
  \comment{onlineaddress}http://xxx.lanl.gov/abs/gr-qc/9805023.
  \keywords{Constraint equations, Black holes, Binary systems}

\bibitem{misner-1963}\comment{article}
\comment{author}Misner, C.W., \comment{title}``The Method of Images in
  Geometrostatics'', \comment{journal}{\em Ann. Phys. (N. Y.)},
  \comment{volume}{\bf 24}, \comment{pages}102--117, (\comment{month}October,
  \comment{year}1963). \keywords{Initial value problem, Black holes}

\bibitem{MTW}\comment{book}
\comment{author}Misner, C.W., Thorne, K.S., and Wheeler, J.A.,
  \comment{title}{\em Gravitation}, (\comment{publisher}W. H. Freeman and
  Company, \comment{address}New York, New York, \comment{year}1973).
  \keywords{Exact solutions, Black Holes, Kerr-Newman metric}

\bibitem{omurchada-york-1973}\comment{article}
\comment{author}{\'O}~Murchadha, N., and York, Jr., J.W.,
  \comment{title}``Existence and uniqueness of solutions of the {H}amiltonian
  constraint of general relativity on compact manifolds'',
  \comment{journal}{\em J. Math. Phys.}, \comment{volume}{\bf
  14}\comment{number}(11), \comment{pages}1551--1557, (\comment{month}November,
  \comment{year}1973). \keywords{Initial value problem, Constraint equations,
  Existence theorems}

\bibitem{omurchada-york-1974-I}\comment{article}
\comment{author}{\'O}~Murchadha, N., and York, Jr., J.W.,
  \comment{title}``Initial-value problem of general relativity. {I}. {G}eneral
  formulation and physical interpretation'', \comment{journal}{\em Phys. Rev.
  D}, \comment{volume}{\bf 10}\comment{number}(2), \comment{pages}428--436,
  (\comment{month}July, \comment{year}1974). \keywords{Initial value problem,
  Constraint equations}

\bibitem{omurchada-york-1974-II}\comment{article}
\comment{author}{\'O}~Murchadha, N., and York, Jr., J.W.,
  \comment{title}``Initial-value problem of general relativity. {II}.
  {S}tability of solution of the initial-value equations'',
  \comment{journal}{\em Phys. Rev. D}, \comment{volume}{\bf
  10}\comment{number}(2), \comment{pages}437--446, (\comment{month}July,
  \comment{year}1974). \keywords{Initial value problem, Constraint equations,
  Stability theory}

\bibitem{omurchada-york-1976}\comment{article}
\comment{author}{\'O}~Murchadha, N., and York, Jr., J.W.,
  \comment{title}``Gravitational Potentials: {A} Constructive Approach to
  General Relativity'', \comment{journal}{\em Gen. Relativ. Gravit.},
  \comment{volume}{\bf 7}\comment{number}(3), \comment{pages}257--261,
  (\comment{month}March, \comment{year}1976). \keywords{Initial value problem,
  Constraint equations}

\bibitem{oppenheimer-volkoff-1939}\comment{article}
\comment{author}Oppenheimer, J.R., and Volkoff, G., \comment{title}``On massive
  neutron cores'', \comment{journal}{\em Phys. Rev.}, \comment{volume}{\bf
  55}\comment{number}(4), \comment{pages}374--381, (\comment{month}February,
  \comment{year}1939). \keywords{Neutron stars}

\bibitem{painleve-1921}\comment{article}
\comment{author}Painlev{\'e}, P., \comment{journal}{\em C. R. Acad. Sci.},
  \comment{volume}{\bf 173}, \comment{pages}677, (\comment{year}1921).
  \keywords{Exact solutions, Black holes, Schwarzschild metric}

\bibitem{pfeiffer-etal-2000}\comment{online}
\comment{author}Pfeiffer, H.P., Teukolsky, S.A., and Cook, G.B.,
  \comment{onlinetitle}``Quasi-circular Orbits for Spinning Binary Black
  Holes'', (\comment{onlinemonth}June, \comment{onlineyear}2000),
  \comment{fileformat}[Online Los Alamos Archive Preprint]: cited on
  \comment{cited}18 July 2000,
  \comment{onlineaddress}http://xxx.lanl.gov/abs/gr-qc/0006084. Submitted to
  {\it Phys. Rev. D}. \keywords{Numerical relativity, Binary systems, Black
  holes}

\bibitem{rieth-Math-Grav}\comment{inbook}
\comment{author}Rieth, R., \comment{title}``On the validity of {W}ilson's
  approach to general relativity'', in \comment{editor}Kr{\'o}lak, A., ed.,
  \comment{booktitle}{\em Mathematics of Gravitation. {P}art {II}.
  {G}ravitational Wave Detection}, \comment{pages} 71--74,
  (\comment{publisher}Polish Academy of Sciences, Institute of Mathematics,
  \comment{address}Warsaw, \comment{year}1997). Proceedings of the Workshop on
  Mathematical Aspects of Theories of Gravitation held in Warsaw, February
  29--March 30, 1996. \keywords{Approximation methods, Binary systems}

\bibitem{robertson-noonan-1968}\comment{book}
\comment{author}Robertson, H.P., and Noonan, T.W., \comment{title}{\em
  Relativity and Cosmology}, (\comment{publisher}Saunders,
  \comment{address}London, \comment{year}1968). \keywords{Exact solutions,
  Black holes, Schwarzschild metric}

\bibitem{shapiro-teukolsky-1980}\comment{article}
\comment{author}Shapiro, S.L., and Teukolsky, S.A.,
  \comment{title}``Gravitational Collapse to Neutron Stars and Black Holes:
  {C}omputer Generation of Spherical Spacetimes'', \comment{journal}{\em
  Astrophys. J.}, \comment{volume}{\bf 235}\comment{number}(1),
  \comment{pages}199--215, (\comment{month}January, \comment{year}1980).
  \keywords{Gravitational collapse, Neutron stars, Black holes}

\bibitem{shapiro-teukolsky-1992}\comment{article}
\comment{author}Shapiro, S.L., and Teukolsky, S.A.,
  \comment{title}``Gravitational collapse of rotating spheroids and the
  formation of naked singularities'', \comment{journal}{\em Phys. Rev. D},
  \comment{volume}{\bf 45}\comment{number}(6), \comment{pages}2006--2012,
  (\comment{month}March, \comment{year}1992). \keywords{Gravitational collapse,
  Black holes}

\bibitem{shibata-1998}\comment{article}
\comment{author}Shibata, M., \comment{title}``A relativistic formalism for
  computation of irrotational binary stars in quasiequilibrium states'',
  \comment{journal}{\em Phys. Rev. D}, \comment{volume}{\bf
  58}\comment{number}(2), \comment{pages}024012/1--5, (\comment{month}July,
  \comment{year}1998). For a related online version see:
  \comment{author}M.~Shibata, \comment{onlinetitle}``A relativistic formalism
  for computation of irrotational binary stars in quasi-equilibrium states'',
  (\comment{onlinemonth}March, \comment{onlineyear}1998),
  \comment{fileformat}[Online Los Alamos Archive Preprint]: cited on
  \comment{cited}18 July 2000,
  \comment{onlineaddress}http://xxx.lanl.gov/abs/gr-qc/9803085.
  \keywords{Neutron stars, Binary systems, Hydrodynamics}

\bibitem{smarr-etal-1976}\comment{article}
\comment{author}Smarr, L., {\v C}ade{\v z}, A., DeWitt, B., and Eppley, K.,
  \comment{title}``Collision of two black holes: {T}heoretical framework'',
  \comment{journal}{\em Phys. Rev. D}, \comment{volume}{\bf
  14}\comment{number}(10), \comment{pages}2443--2452, (\comment{month}November,
  \comment{year}1976). \keywords{Numerical relativity, Initial value problem,
  Black holes}

\bibitem{smarr-york-1978}\comment{article}
\comment{author}Smarr, L., and York, Jr., J.W., \comment{title}``Kinematical
  conditions in the construction of spacetime'', \comment{journal}{\em Phys.
  Rev. D}, \comment{volume}{\bf 17}\comment{number}(10),
  \comment{pages}2529--2551, (\comment{month}May, \comment{year}1978).
  \keywords{ADM formalism, Cauchy problem, Gauge conditions}

\bibitem{sorkin-1981}\comment{article}
\comment{author}Sorkin, R.D., \comment{title}``A Criterion for the Onset of
  Instabilities at a Turning Point'', \comment{journal}{\em Astrophys. J.},
  \comment{volume}{\bf 249}\comment{number}(1), \comment{pages}254--257,
  (\comment{month}October, \comment{year}1981). \keywords{Stability theory}

\bibitem{sorkin-1982}\comment{article}
\comment{author}Sorkin, R.D., \comment{title}``A Stability Criterion for
  Many-Parameter Equilibrium Families'', \comment{journal}{\em Astrophys. J.},
  \comment{volume}{\bf 257}\comment{number}(2), \comment{pages}847--854,
  (\comment{month}June, \comment{year}1982). \keywords{Stability theory}

\bibitem{stergioulas-1998}\comment{online}
\comment{author}Stergioulas, N., \comment{onlinetitle}``Rotating Stars in
  Relativity'', (\comment{onlinemonth}June, \comment{onlineyear}1998),
  \comment{fileformat}[Article in Online Journal Living Reviews in Relativity]:
  cited on \comment{cited}18 July 2000,
  \comment{onlineaddress}http://www.livingreviews.org/Articles/Volume1/1998-8s%
tergio. \keywords{Relativistic stars, Perturbation theory}

\bibitem{stergioulas-friedman-1995}\comment{article}
\comment{author}Stergioulas, N., and Friedman, J.L., \comment{title}``Comparing
  Models of Rapidly Rotating Relativistic Stars Constructed by Two Numerical
  Methods'', \comment{journal}{\em Astrophys. J.}, \comment{volume}{\bf
  444}\comment{number}(1), \comment{pages}306--311, (\comment{month}May,
  \comment{year}1995). For a related online version see:
  \comment{author}N.~Stergioulas, et al., \comment{onlinetitle}``Comparing
  Models of Rapidly Rotating Relativistic Stars Constructed by Two Numerical
  Methods'', (\comment{onlinemonth}November, \comment{onlineyear}1994),
  \comment{fileformat}[Online Los Alamos Archive Preprint]: cited on
  \comment{cited}18 July 2000,
  \comment{onlineaddress}http://xxx.lanl.gov/abs/astro-ph/9411032.
  \keywords{Neutron stars, Numerical methods}

\bibitem{teukolsky-1998}\comment{article}
\comment{author}Teukolsky, S.A., \comment{title}``Irrotational Binary Neutron
  Stars in Quasiequilibrium in General Relativity'', \comment{journal}{\em
  Astrophys. J.}, \comment{volume}{\bf 504}\comment{number}(1),
  \comment{pages}442--449, (\comment{month}September, \comment{year}1998). For
  a related online version see: \comment{author}S.A. Teukolsky,
  \comment{onlinetitle}``Irrotational Binary Neutron Stars in Quasiequilibrium
  in General Relativity'', (\comment{onlinemonth}March,
  \comment{onlineyear}1998), \comment{fileformat}[Online Los Alamos Archive
  Preprint]: cited on \comment{cited}18 July 2000,
  \comment{onlineaddress}http://xxx.lanl.gov/abs/gr-qc/9803082.
  \keywords{Neutron stars, Binary systems, Hydrodynamics}

\bibitem{thornburg-1987}\comment{article}
\comment{author}Thornburg, J., \comment{title}``Coordinate and boundary
  conditions for the general relativistic initial data problem'',
  \comment{journal}{\em Class. Quantum Grav.}, \comment{volume}{\bf
  4}\comment{number}(5), \comment{pages}1119--1131, (\comment{month}September,
  \comment{year}1987). \keywords{Initial value problem, Constraint equations,
  Black holes}

\bibitem{Essays-in-GR}\comment{proceedings}
\comment{editor}Tipler, F.J., ed., \comment{title}{\em Essays in General
  Relativity}, (\comment{publisher}Academic, \comment{address}New York,
  \comment{year}1980). \keywords{}

\bibitem{uryu-eriguchi-2000}\comment{article}
\comment{author}Ury{\=u}, K., and Eriguchi, Y., \comment{title}``New numerical
  method for constructing quasiequilibrium sequences of irrotational binary
  neutron stars in general relativity'', \comment{journal}{\em Phys. Rev. D},
  \comment{volume}{\bf 61}\comment{number}(12), \comment{pages}124023/1--19,
  (\comment{month}June, \comment{year}2000). For a related online version see:
  \comment{author}K.~Ury{\=u}, et al., \comment{onlinetitle}``A new numerical
  method for constructing quasi-equilibrium sequences of irrotational binary
  neutron stars in general relativity'', (\comment{onlinemonth}August,
  \comment{onlineyear}1999), \comment{fileformat}[Online Los Alamos Archive
  Preprint]: cited on \comment{cited}18 July 2000,
  \comment{onlineaddress}http://xxx.lanl.gov/abs/gr-qc/9908059.
  \keywords{Neutron stars, Binary systems, Numerical methods}

\bibitem{wilson-1972}\comment{article}
\comment{author}Wilson, J.R., \comment{title}``Models of Differentially
  Rotating Stars'', \comment{journal}{\em Astrophys. J.}, \comment{volume}{\bf
  176}\comment{number}(1), \comment{pages}273--286, (\comment{month}August,
  \comment{year}1972). \keywords{Neutron stars, Numerical methods}

\bibitem{wilson-mathews-Frontiers}\comment{inbook}
\comment{author}Wilson, J.R., and Mathews, G.J., \comment{title}``Relativistic
  Hydrodynamics'', \comment{pages} 306--314, In Evans et~al. \cite{Frontiers}.
  \keywords{Neutron stars, Hydrodynamics, Approximation methods}

\bibitem{wilson-mathews-1995}\comment{article}
\comment{author}Wilson, J.R., and Mathews, G.J., \comment{title}``Instabilities
  in close neutron star binaries'', \comment{journal}{\em Phys. Rev. Lett.},
  \comment{volume}{\bf 75}\comment{number}(23), \comment{pages}4161--4164,
  (\comment{month}December, \comment{year}1995). \keywords{Neutron stars,
  Binary systems, Gravitational collapse, Approximation methods}

\bibitem{wilson-etal-1996}\comment{article}
\comment{author}Wilson, J.R., Mathews, G.J., and Marronetti, P.,
  \comment{title}``Relativistic Numerical Method for Close Neutron Star
  Binaries'', \comment{journal}{\em Phys. Rev. D}, \comment{volume}{\bf
  54}\comment{number}(2), \comment{pages}1317--1331, (\comment{month}July,
  \comment{year}1996). For a related online version see: \comment{author}J.R.
  Wilson, et al., \comment{onlinetitle}``Relativistic Numerical Method for
  Close Neutron Star Binaries'', (\comment{onlinemonth}January,
  \comment{onlineyear}1996), \comment{fileformat}[Online Los Alamos Archive
  Preprint]: cited on \comment{cited}18 July 2000,
  \comment{onlineaddress}http://xxx.lanl.gov/abs/gr-qc/9601017.
  \keywords{Neutron stars, Binary systems, Numerical methods}

\bibitem{york-1971}\comment{article}
\comment{author}York, Jr., J.W., \comment{title}``Gravitational degrees of
  freedom and the initial-value problem'', \comment{journal}{\em Phys. Rev.
  Lett.}, \comment{volume}{\bf 26}\comment{number}(26),
  \comment{pages}1656--1658, (\comment{month}June, \comment{year}1971).
  \keywords{Cauchy problem, Initial value problem}

\bibitem{york-1972}\comment{article}
\comment{author}York, Jr., J.W., \comment{title}``Role of conformal
  three-geometry in the dynamics of gravitation'', \comment{journal}{\em Phys.
  Rev. Lett.}, \comment{volume}{\bf 28}\comment{number}(16),
  \comment{pages}1082--1085, (\comment{month}April, \comment{year}1972).
  \keywords{Cauchy problem, Initial value problem}

\bibitem{york-1973}\comment{article}
\comment{author}York, Jr., J.W., \comment{title}``Conformally invariant
  orthogonal decomposition of symmetric tensors on {R}iemannian manifolds and
  the initial-value problem of general relativity'', \comment{journal}{\em J.
  Math. Phys.}, \comment{volume}{\bf 14}\comment{number}(4),
  \comment{pages}456--464, (\comment{month}April, \comment{year}1973).
  \keywords{Cauchy problem, Initial value problem, Constraint equations}

\bibitem{york-1974}\comment{article}
\comment{author}York, Jr., J.W., \comment{title}``Covariant decompositions of
  symmetric tensors in the theory of gravitation'', \comment{journal}{\em Ann.
  Inst. Henri Poincar{\'e}, A}, \comment{volume}{\bf 21}\comment{number}(4),
  \comment{pages}319--332, (\comment{year}1974). \keywords{Cauchy problem,
  Initial value problem, Constraint equations}

\bibitem{york-Sources}\comment{inbook}
\comment{author}York, Jr., J.W., \comment{title}``Kinematics and Dynamics of
  General Relativity'', in \comment{editor}Smarr, L.L., ed.,
  \comment{booktitle}{\em Sources of Gravitational Radiation}, \comment{pages}
  83--126, (\comment{publisher}Cambridge University Press,
  \comment{address}Cambridge, England, \comment{year}1979). \keywords{ADM
  formalism, Cauchy problem}

\bibitem{york-Essays-GR}\comment{inbook}
\comment{author}York, Jr., J.W., \comment{title}``Energy and Momentum of the
  Gravitational Field'', \comment{pages} 39--58, In Tipler \cite{Essays-in-GR}.
  \keywords{Energy and momentum, Asymptotics, Gauge conditions, Black holes}

\bibitem{york-1984}\comment{article}
\comment{author}York, Jr., J.W., \comment{title}``Initial data for {$N$} black
  holes'', \comment{journal}{\em Physica (Utrecht)}, \comment{volume}{\bf
  124A}\comment{number}(1--3), \comment{pages}629--637, (\comment{month}March,
  \comment{year}1984). \keywords{Initial value problem, Constraint equations,
  Black holes}

\bibitem{york-Frontiers}\comment{inbook}
\comment{author}York, Jr., J.W., \comment{title}``Initial Data for Collisions
  of Black Holes and Other Gravitational Miscellany'', \comment{pages} 89--109,
  In Evans et~al. \cite{Frontiers}. \keywords{Initial value problem, Constraint
  equations, Black holes}

\bibitem{york-1999}\comment{article}
\comment{author}York, Jr., J.W., \comment{title}``Conformal `thin-sandwich'
  data for the initial-value problem of general relativity'',
  \comment{journal}{\em Phys. Rev. Lett.}, \comment{volume}{\bf
  82}\comment{number}(7), \comment{pages}1350--1353, (\comment{month}February,
  \comment{year}1999). For a related online version see: \comment{author}J.W.
  York, Jr., \comment{onlinetitle}``Conformal `thin-sandwich' data for the
  initial-value problem of general relativity'', (\comment{onlinemonth}October,
  \comment{onlineyear}1998), \comment{fileformat}[Online Los Alamos Archive
  Preprint]: cited on \comment{cited}18 July 2000,
  \comment{onlineaddress}http://xxx.lanl.gov/abs/gr-qc/9810051.
  \keywords{Initial value problem, Constraint equations}

\bibitem{york-piran-StG}\comment{inbook}
\comment{author}York, Jr., J.W., and Piran, T., \comment{title}``The Initial
  Value Problem and Beyond'', in \comment{editor}Matzner, R.A., and Shepley,
  L.C., eds., \comment{booktitle}{\em Spacetime and Geometry}, \comment{pages}
  147--176, (\comment{publisher}University of Texas, \comment{address}Austin,
  \comment{year}1982). \keywords{Initial value problem, Constraint equations,
  Finite difference methods, Black holes}

\end{thebibliography}
\end{document}